\documentclass{article}

\usepackage{arxiv}

\usepackage{graphicx}% Include figure files
\usepackage{dcolumn}% Align table columns on decimal point
\usepackage{listings}

\usepackage{bm}% bold math
\usepackage[mathlines]{lineno}% Enable numbering of text and display math
\usepackage{makecell}
\usepackage{multirow}
\usepackage{hhline}
\usepackage[utf8]{inputenc}
\usepackage[T1]{fontenc}
\usepackage{mathptmx}
\usepackage{etoolbox}
\usepackage{xcolor}
\usepackage{color}
\usepackage{float}
\usepackage{hyperref}
\newcommand{\be}{\begin{equation}}
\newcommand{\ee}{\end{equation}}

\newcommand\numberthis{\addtocounter{equation}{1}\tag{\theequation}}

\usepackage{amsmath}
\usepackage{natbib}

\usepackage[utf8]{inputenc} % allow utf-8 input
\usepackage[T1]{fontenc}    % use 8-bit T1 fonts
\usepackage{hyperref}       % hyperlinks
\usepackage{url}            % simple URL typesetting
\usepackage{booktabs}       % professional-quality tables
\usepackage{amsfonts}       % blackboard math symbols
\usepackage{nicefrac}       % compact symbols for 1/2, etc.
\usepackage{microtype}      % microtypography
\usepackage{lipsum}
\usepackage{graphicx}
\graphicspath{ {./images/} }

\title{Wind Farm Dynamics over a Diurnal Cycle: Analysis of a Comprehensive Large Eddy Simulation, Web-Services Accessible Dataset}

\author{
 Shuolin Xiao \\
  Ralph O’Connor Sustainable Energy Institute \\ Johns Hopkins University\\ Baltimore, MD 21218, USA. \\
  \texttt{} \\
   \And
 Xiaowei Zhu \\
  Department of Mechanical and Materials Engineering \\ Portland State University \\ Portland 97201, OR, USA. \\
  \texttt{} \\
  \And
 Ghanesh Narasimhan \\
  Department of Mechanical Engineering \\\& St. Anthony Falls Lab.\\ University of Minnesota \\ Minneapolis, MN 55455, USA. \\
  \texttt{} \\
  \And
 Dennice F Gayme \\
  Department of Mechanical Engineering \\ Johns Hopkins University \\ Baltimore, MD 21218, USA. \\
  \texttt{} \\
  \And
 Charles Meneveau \\
  Department of Mechanical Engineering \\ Johns Hopkins University \\ Baltimore, MD 21218, USA. \\ \texttt{meneveau@jhu.edu}
}

\begin{document}
\maketitle
\begin{abstract}
The atmospheric boundary layer undergoes significant changes throughout a diurnal cycle, affecting wind turbine performance and wakes in wind farms. Wind farm Large Eddy Simulations (LES) under such conditions provide rich datasets to study the underlying dynamics and identify important trends. Here, we describe a comprehensive open dataset generated using LES of an 8-turbine wind farm consisting of \textcolor{black}{four rows of two turbines.} To avoid specifying either prescribed surface temperature or heat flux, a local 1D soil heat conduction model is used with time-periodic solar surface heating, coupled to  LES.  After several days of low-resolution LES, an approximately time periodic behavior is achieved, after which high-resolution LES is continued during a 24-hour period. Analysis of the LES data reveals that wind turbine wakes have a significant impact on the temperature field and spatial surface heat flux patterns and exhibiting increased surface temperature behind the wind farm at night under the specific conditions of the simulation (dry unvegetated soil, clear sky). We observe that for a few morning hours the first row of wind turbines generates less power compared to the last row. Detailed analyses of the data using innovative web-services facilitated data access tools reveal that during the morning transition, the \textcolor{black}{presence} of a low-level jet and the wind farm blockage effect combine to cause cooling and a reduction in wind speed at hub height upstream of the wind farm. In addition, larger turbulence levels exist downstream in the wind farm, explaining the larger power production of downstream turbines. 
\end{abstract}

% keywords can be removed

\section{Introduction}

The wind and temperature fields in the atmospheric boundary layer (ABL), which influence the background stratification and consequently the atmospheric conditions in the ABL, are the most critical meteorological variables for wind energy applications. A number of numerical studies have investigated the characteristics of the convective ABL \cite{mason1989large,Liu_et_al_AMS_2023}, conventionally neutral ABL \cite{ayotte1996evaluation,Liu_Stevens_PNAS_2022}, and stable ABL \cite{mason1990large,stoll2020large,narasimhanBLM2024,Shen_Liu_Lu_Stevens_2024}, typically conducted with constant thermal forcing at the ground surface and analyzing the statistics when quasi-equilibrium conditions are reached. The stability of the ABL flow affects wind shear, wind veer, and turbulence in the incoming wind profile, all of which are essential to wind farm performance \cite{narasimhan2022effects,narasimhan2024analytical,narasimhan2024extendedanalyticalwakemodel,narasimhanBLM2024,kelly_laan-2023}. Furthermore, recent studies have examined the interaction between the aforementioned ABL flow and wind farms \cite{porte2011large,allaerts2015large}, revealing that the presence of wind farms generates wake flows, triggers turbulence behind the wind turbines, and even affects the temperature field distribution within and beyond the wind farm \cite{lu2011large,strickland2022wind}. In particular, nighttime warming behind wind farms has been reported in a few high-fidelity numerical studies on the interaction between a stably stratified ABL and a wind farm. This temperature increase has also been confirmed by several observational studies, which show a rise of approximately $O(1)$ K at the land surface, depending on vegetation type and season \cite{zhou2012impacts,rajewski2013crop,smith2013situ}. 

On a daily scale, the real ABL flow not only experiences stable, conventionally neutral, or convective conditions but also undergoes late-afternoon and early-morning transitions, during which stability and thermal forcing at the ground surface change significantly within a few hours—not only in magnitude but also in sign—substantially influencing the ABL flow and temperature structure. In the diurnal cycle, variations in the atmospheric boundary layer height lead to changes in wind direction in part due to the Coriolis effect. Additionally, the early-morning 
transitions \textcolor{black}{can trigger formation of a} low-level jet, a narrow band of fast-moving wind flow above the ground surface which enhances wind speeds near the ground.
%and contributes to increased turbulence and mixing in the lower atmosphere. 
There are a few large-eddy simulation (LES) studies of the ABL and its interaction with wind farms on a daily cycle time scale. Several LES studies investigated the evolution of ABL flows and the temperature field during a typical diurnal cycle, using surface temperature or surface sensible heat flux from a well-known field campaign as thermal boundary conditions \cite{kumar2006large,kumar2010impact}. Quon {\it et al.} examined the interaction of ABL flows with a wind farm using realistic turbine spacing and a wake-steering strategy that varies between different turbines, all on a diurnal scale through measurement-driven large-eddy simulations \cite{quon2024measurement}. Limited studies explored wind farm performance in a diurnal cycle through an idealized LES, where the formation of a low-level jet and its modulation by the presence of the wind farm were observed in the early morning \cite{abkar2016wake,sharma2017perturbations}. 

The presence of wind turbines will generate wake flows behind them, causing significant reductions in power for turbines in the wakes of upstream turbines \cite{frandsen1992wind,stevens2017flow, porte2020wind}. In addition, the wake turbulence is expected to modulate the ground surface sensible heat flux due to the air flow above it. Therefore, ground surface sensible heat flux, and consequently ground surface temperature, are expected to be heterogeneous between the areas directly behind the wind turbines and the spaces between them. Most diurnal cycle studies impose either spatially uniform surface temperature or surface sensible heat flux from the field campaign as thermal boundary conditions \cite{kumar2006large,kumar2010impact,cortina2017turbulence,allaerts2023using}. More realistically, the ground surface sensible heat flux and temperature can be determined by coupling the solution of a three-dimensional heat conduction equation in the soil with a surface thermal energy balance approach. \cite{lu2015impact,abkar2016wake}. The soil model can also be simplified to a zero-layer model, without explicitly resolving the soil domain \cite{van2019idealized}.

In this study, we investigate a representative diurnal cycle evolution of ABL flow and its interactions with a small wind farm. 
We aim to answer the following research questions: (a) What is the effect of the wind farm on the thermal state of the ABL especially in the wake of the wind farm during the stably stratified conditions during night-time? (b) Is the usual turbine wake effect (namely that downstream turbines can generate substantially less power than upstream turbines when they are waked) observed at all times during a representative 24 hour period of a typical wind farm evolution?  (c) Can carefully designed storage of the simulation data facilitate the identification of  physical phenomena that underlie the observations made? The amount of data generated during a diurnal cycle of a wind-turbine array boundary layer flow (WTABL) is quite significant and provides a rich amount of information that could be used to address also many other research questions. Therefore, both to facilitate data analysis to answer the aforementioned current research questions and to enable additional analysis of many other possible questions, the dataset generated in our simulation is ingested into an existing  public database system (the Johns Hopkins Database System, JHTDB). Together with another wind farm dataset for a larger wind farm in a quasi-steady state conventionally neutral ABL \cite{Zhu2025jhtdbwindpaper}, the new datasets forms the new JHTDB-wind system (\url{https://turbulence.idies.jhu.edu/datasets/windfarms}). 

The LES framework and governing equations are described in section \ref{sec:LESdescription}, including a description of the boundary and inflow conditions, the latter using a concurrent-precursor inflow method \cite{stevens2014concurrent}. For the thermal boundary condition, a separate one-dimensional soil heat conduction equation is used to avoid prescribing homogeneous ground surface temperature or heat fluxes a priori. Section \ref{sec:simulationdata} provides relevant information regarding the simulation setup and data stored to build the JHTDB-wind database. Simulation results and general observations regarding the flow evolution during the 24 hour period are presented and discussed in section \ref{sec:generalobs}. In section \ref{sec:resultstemp} and \ref{sec:resultspower} we discuss results pertaining to the two specific research questions regarding the thermal state and power production during transitions to and from the stably stratified conditions. Conclusions are provided in \S \ref{sec:conclusions}. Details about JHTDB-wind as well as descriptions of methods to access the public dataset are provided in appendices.  

\section{Large-eddy simulation}
\label{sec:LESdescription}
Simulations are performed using the open-source large-eddy simulation model LESGO. This code has been  applied extensively \cite{calaf2010large,stevens2017flow, martinez2017large,stevens2018comparison, gharaati2022large, shapiro2018,shapiro2020, narasimhan2022effects,
narasimhan2024analytical,
narasimhan2024extendedanalyticalwakemodel, gharaati2024large} in previous LES studies to simulate wind turbine wake flows. The three-dimensional Cartesian coordinates are $(x, y, z)$, where $x$ is the streamwise direction, $y$ the spanwise direction, and $z$ the vertical direction, respectively. We also use, interchangeably when \textcolor{black}{convenient},  index notation where $x_i$ ($i$ = 1, 2, 3) for each direction. Velocity components along these directions are  $u_i$ ($i$ = 1, 2, 3), or $u$, $v$, and $w$ for the $x$, $y$, and $z$-directions, respectively.

The turbulent flow is simulated by solving the filtered Navier-Stokes equations in their rotational form assuming divergence-free velocity distribution and in the Boussinesq approximation, along with the transport equation for the potential temperature field:
% \begin{equation}
%     \frac{\partial \tilde u_i}{\partial x_i} = 0,
% \end{equation} 
%\begin{equation}
\begin{align*}
    \frac{\partial \tilde u_i}{\partial x_i} &= 0,\numberthis\label{eq:continuity}\\
        \frac{\partial \tilde u_i}{\partial t} 
        + \tilde u_j \left( \frac{\partial \tilde u_i}{\partial x_j} - \frac{\partial \tilde u_j}{\partial x_i} \right) 
        &= -\frac{\partial \tilde p^{*}}{\partial x_i} 
        + \frac{g}{\theta_0} (\tilde \theta - \tilde \theta_0)\delta_{i3} 
        %+ \nu \frac{\partial^2 \tilde u_i}{\partial x_j^2} 
        - \frac{\partial \tau^{\mathrm{SGS},d}_{ij}}{\partial x_j} 
        - f_i 
        \\
        &\mspace{20mu}+ f_c(\tilde u_2 - V_g) \delta_{i1} - f_c(\tilde u_1 - U_g) \delta_{i2},\numberthis\label{eq:momentumtransport}\\
         \frac{\partial \tilde \theta}{\partial t} +\tilde u_j\frac{\partial \tilde \theta}{\partial x_j} &= -\frac{\partial \Pi_j}{\partial x_j}.\numberthis
    \label{eq:temperaturetransport}
\end{align*}
%\end{equation}
% and,
% \begin{equation}
%     \frac{\partial \tilde \theta}{\partial t} +\tilde u_j\frac{\partial \tilde \theta}{\partial x_j} = -\frac{\partial \Pi_j}{\partial x_j}.
%     \label{eq:temperaturetransport}
% \end{equation}
Tilde ($\tilde\cdot$) notation indicate filtering at the LES grid scale $\tilde{\Delta}=\sqrt[3]{\Delta x \ \Delta y \ \Delta z}$. Also, $\rho$ is air density; $\tau^{\text{SGS}}_{ij}$ is the subgrid-scale (SGS) stress tensor and $\tau^{\text{SGS},d}_{ij}=\tau^{\text{SGS}}_{ij}-(1/3)\tau^{\text{SGS}}_{kk} \delta_{ij}$ is the deviatoric (trace-free) part
of the SGS stress tensor, $\tilde p^{*} = \tilde p/\rho+\tilde u_k\tilde u_k/2+\tau^{\text{SGS}}_{kk}/3$ is the pseudo pressure, where $\tilde p$ is the resolved   pressure; while $f_i$ is the  body force that models the  aerodynamic forces exerted on the air flow by the turbines. Here $f_i$ is modeled using the actuator disk model (ADM) during the initial   coarse simulation and the actuator line model (ALM) in the fine simulation (more details provided in \ref{sec:wind turbine}). Moreover  $g = 9.81\,\text{m}/\text{s}^2$ is the gravitational acceleration, $\tilde \theta_0$ is the reference potential temperature scale taken to be 263.5 K. The SGS stress tensor $\tau^{\text{SGS},d}_{ij}$ is modeled using the Lilly-Smagorinsky eddy-viscosity  model, i.e. $\tau^{\text{SGS},d}_{ij}=-2\nu_{\text{SGS}}\tilde S_{ij}=-2(C_s\tilde{\Delta})^2|\tilde S|\tilde S_{ij}$, where $\tilde S_{ij}=0.5(\partial \tilde{u}_i/\partial x_j+\partial \tilde{u}_j/\partial x_i)$ is the resolved strain-rate tensor, $|\tilde S|=\sqrt{2\tilde S_{ij}\tilde S_{ij}}$ is the strain-rate magnitude, and $\nu_{\text{SGS}}=(C_s\tilde{\Delta})^2|\tilde S|$ is the  SGS eddy viscosity.  $C_s$ is obtained dynamically utilizing the Lagrangian-averaged scale-dependent dynamic (LASD) model \citep{bou2005scale}. This particular SGS model has been  applied extensively in several prior LES studies of wind turbine wake flows \citep{calaf2010large,stevens2017flow, martinez2017large,stevens2018comparison,
narasimhan2022effects, gharaati2022large, narasimhan2024analytical, gharaati2024large}. In Eq. \ref{eq:temperaturetransport},   $\Pi_j=\widetilde{u_j\theta}-\tilde u_j\tilde \theta$ represents the SGS heat flux in which the eddy diffusivity 
($\kappa_{\text{SGS}}$) is computed as  
$\kappa_{\text{SGS}}=Pr_{\text{SGS}}^{-1}\nu_{\text{SGS}}$, where the SGS Prandtl number is chosen as unity ($Pr_{\text{SGS}}=1$) \citep{narasimhan2022effects}.

LESGO is based on the pseudo-spectral method  for the spatial discretizations in the $x$- and $y$-directions, while a centered, second-order finite-differencing method based on staggered grids is used for the spatial discretization in the $z$-(vertical) direction.  Time advancement uses the second-order Adams-Bashforth scheme to obtain a predicted velocity field, followed by a pressure Poisson equation solution to impose the divergence-free constraint. Then, the velocity field is projected onto the divergence-free space using the gradient of the pseudo-pressure to obtain the velocity field for the new time step. The equations, numerical treatment and boundary conditions are described in additional detail in Ref. \cite{Zhu2025jhtdbwindpaper} and earlier references \cite{albertson1996large, albertson1999surface,stevens2018comparison,narasimhan2022effects}.  
The atmospheric boundary layer flow is driven by a constant Geostrophic wind. In the Coriolis terms, $f_c = 2\Omega\sin\phi = 10^{-4}\ \text{s}^{-1}$ is the Coriolis parameter for a mid-latitude condition. The  geostrophic wind $\mathbf{G} = (U_g, V_g)$ is an imposed parameter.  
 
During the initial stages of the simulation using a coarse grid (see the next section for details), we use the actuator disk representation of the turbine\cite{calaf2010large}. After switching the simulation to finer resolution, the actuator line model (ALM)\cite{sorensen2002numerical,troldborg2009actuator,martinez2015large} is adopted to represent each turbine blade by a rotating line, along which forces are applied according to the relative velocity field and the angle of attack. Lift and drag coefficients at various blade locations are obtained from tabulated data. A Gaussian kernel is used to smear the ALM forces onto the LES grid, using a kernel width of $\epsilon$ equal to twice the LES resolution as recommended by \cite{martinez2015large}. We implement the advanced filtered actuator-line method based on filtered line theory \cite{martinez2019filtered} and its generalization \cite{martinez2024generalized}.
As wind turbine, we use the parameters of the NREL-5MW reference wind turbine.  It is three bladed with diameter $D=126\,\text{m}$ and a hub height of $z_h = 90\,\text{m}$. \textcolor{black}{More details about the turbine model are presented in \S \ref{sec:wind turbine}.} 

\subsection{Boundary conditions}

\begin{figure}[H]
    \centering
    \includegraphics[width=\textwidth]{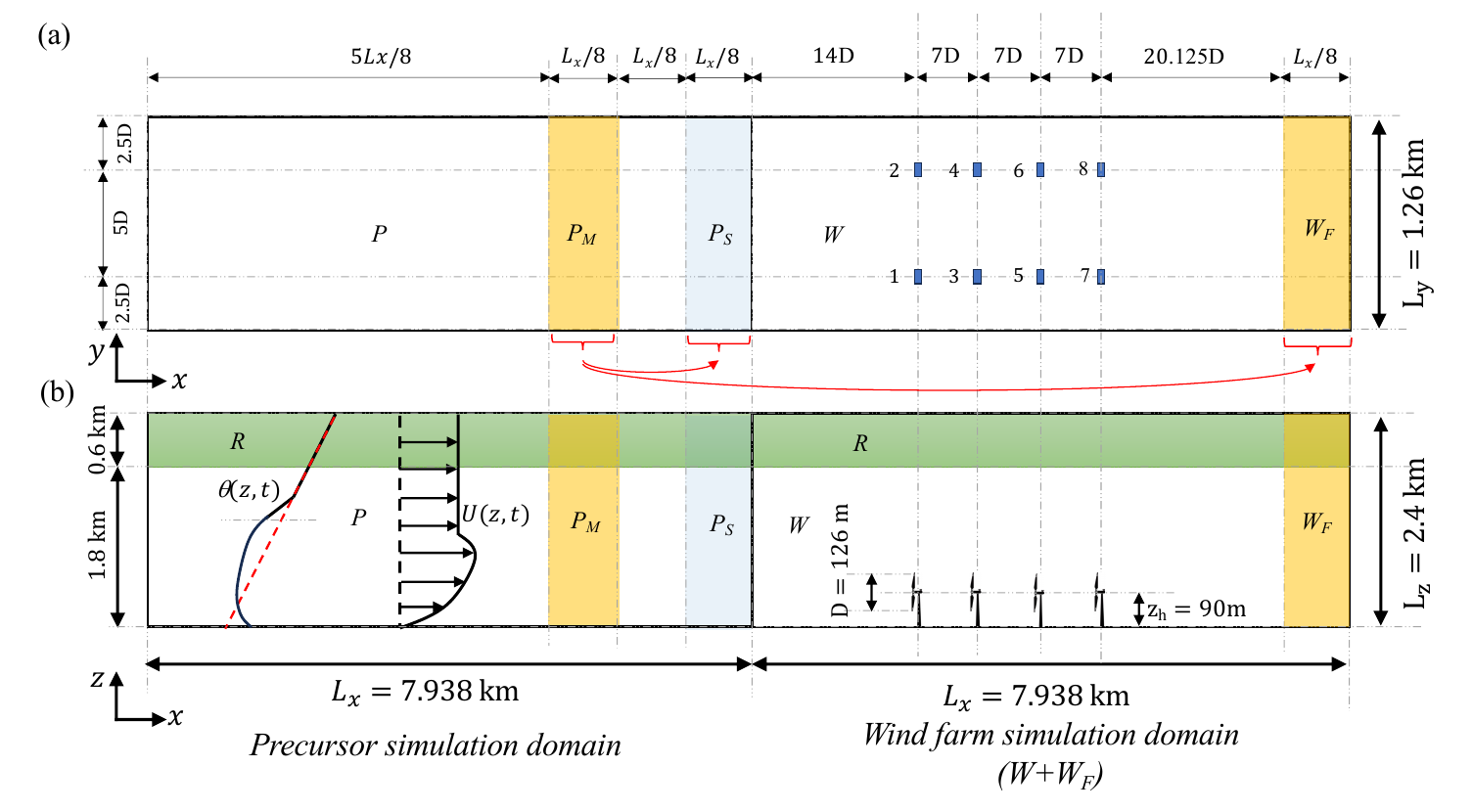}
    \caption{Schematic representation of the computational simulation domain (not to scale), showing: (a) top view ($x$–$y$ plane) and  (b) front view (x-z plane). The precursor computational domain consists of the regions denoted as ``$P$'', the precursor mapping region ``$P_M$'', and the precursor spanwise shifting region ``$P_S$''. The wind farm  computational domain includes the windfarm region ``$W$'' and the fringe region ``$W_F$'' near the outlet. Both precursor and windfarm computational domains include a Rayleigh damping region at the top (region $R$''). The turbine diameter $D = 126$ m and hub height $z_h = 90$ m are also indicated. 
    }
    \label{fig:schematic1}
\end{figure}

Periodic boundary conditions are used in the spanwise $y$ direction, and the stress-free condition is applied at the top  boundary. In the streamwise direction, inflow-outflow conditions together with a concurrent precursor simulation approach \cite{stevens2014concurrent} are used in order to model the inflow. Two simulations are coupled: a precursor domain without wind turbines generates realistic turbulence that is then used in the wind farm domain as inflow. The wind farm domain uses a fringe region to force the outflow to equal the inflow at the periodic boundary (see Fig. \ref{fig:schematic1}). More details of the inflow-outflow conditions implemented in the current pseudo-spectral solver are provided in \cite{stevens2014concurrent}.

Additionally, the precursor domain simulation  uses a fringe region to impose a spanwise shift \cite{munters2016shifted}, which serves to prevent persistent spanwise locking of large-scale turbulent structures. A shift of $L_{y-{\rm shift}} = 761.25\,\text{m}$, which is approximately as $L_{y-{\rm shift}}\approx \frac{1}{4} L_z$, with $L_z$ as the domain height.  The fringe region $W_F$, as well as the mapping ($P_M$) and spanwise shifting ($P_S$) regions, have a length of $L_x/8$, and the mapping region $P_M$ extends from $5L_x/8$ to $3L_x/4$.  Vertically, $1.8\,\text{km}$ is used for unforced data, followed by  a $0.6\,\text{km}$ Rayleigh damping sponge layer ($R$) above (see Figure~\ref{fig:schematic1}). The sponge layer, located at the top 25\% of the domain, includes a damping body force with a cosine profile to suppress the reflection of gravity waves.  A stress-free boundary condition is imposed on the top boundaries of the domains. 

At the bottom surface, we specify Monin-Obukov equilibrium surface fluxes in both the precursor and wind turbine domains using the usual Monin-Obukov \citep{Monin_Obukhov_1954} equilibrium wall modeling, using a prescribed roughness 
length $z_0$ and standard stability correction functions.  The components of local surface shear stress are computed as a function of the prescribed roughness length by

\begin{equation}
    \tau_{\rm {i,3|surf}}=- u_{*}^2\frac{\widehat{\widetilde {u_i}}}{\sqrt{\widehat{\widetilde u}^{2}+\widehat{\widetilde v}^{2}}}, ~~~~~~~ i = 1,2; ~~~~~{\rm and}~~~~~
     u_{*}=\kappa \, \frac{\sqrt{\widehat{\widetilde u}^2(0.5 \Delta z)+\widehat{\widetilde v}^2(0.5 \Delta z)}}{\ln({0.5 \Delta z}/{z_0})-\Psi_m({0.5 \Delta z}/{L})+\Psi_m(z_0/L)}.
\end{equation}
Here, $\kappa = 0.41$ is the von K\'arm\'an constant, $z_0 = 0.1$ m is the prescribed roughness length, the friction velocity $u_*$ is expressed in terms of the horizontal velocity $({\widehat{\widetilde u}},{\widehat{\widetilde v}})$ at the first grid-point ($z_1=0.5 \Delta z)$, filtered at twice the grid resolution, $\hat{\tilde{\Delta}}=2\tilde{\Delta}$ \citep{bou2005scale}.  
Moreover, $\psi_m$  is the momentum stability correction function   for the momentum flux boundary condition which is specified according to MOST \citep{Chenge_Brutsaert_2005} following,
\begin{equation}
    \Psi_m\left(\zeta=\frac{z}{L}\right)=\int_{0}^{\zeta}[1-\phi_m(\zeta^\prime)]\frac{d\zeta^\prime}{\zeta^\prime}\,.  \label{Psi_m_func}
\end{equation}
Here, $\zeta = z/L$ is the stability parameter, $L$ is the Obukhov length (defined below). The stability correction function $\Psi_m(\zeta)$ is obtained by integrating $[1-\phi_m(\zeta)]/\zeta$.  Here, $\phi_m(\zeta)=\kappa z/u_* \partial |u|/\partial z$ is the stability function that physically represents the dimensionless form of velocity gradient ($\partial |u|/\partial z$), made non-dimensional using friction velocity $u_*$ as the characteristic velocity scale and $\kappa z$ as the characteristic mixing length scale of the turbulent eddies within the atmospheric surface layer. We use the  empirical expressions from \cite{Chenge_Brutsaert_2005,Brutsaert_2005} for $\phi_m(\zeta)$ under unstable ($\zeta<0$) and stable ($\zeta>0$) atmospheric conditions.
\textcolor{black}{They are listed in detail in Appendix A.}

The Obukhov length $L$ is computed in LES as, 
\begin{equation}
    L=\frac{u_*^2 \, \tilde\theta(0.5 \Delta z)}{\kappa g T_*},
\end{equation}
where $\tilde\theta(0.5 \Delta z)$ is the potential temperature at the first grid point and $T_*$ potential temperature flux scale. It is determined from the MOST as follows,
\begin{equation}
  T_*=-\frac{ q_{\rm {3|surf}}}{u_*},
\label{eq:air turbulent heat flux}
\end{equation}
where $q_{\rm {3|surf}}$ is the
imposed surface heat flux given as,
\begin{equation}
   q_{\rm {3|surf}}=\kappa \, u_{*} \, \frac{\tilde\theta_{s}-\tilde\theta(0.5 \Delta z)}{\log( 0.5 \Delta z /z_{0,s})-\Psi_h(0.5 \Delta z/L)+\Psi_h(z_{0,s}/L)},
\label{eq:air turbulent heat flux}
\end{equation}
and where $z_{0,s}=0.1 z_0$ \citep{Brutsaert_2005}, $\theta_s$ is the surface potential temperature, $\tilde\theta(0.5 \Delta z) $ is the potential temperature at the first grid point $(z=0.5\Delta z)$, and $\Psi_h(\zeta)$ is the stability  correction function determined from the MOST theory. It is obtained as,
\begin{equation}
\Psi_h\left(\zeta\right)=\int_{0}^{\zeta}[1-\phi_h(\zeta^\prime )]\frac{d\zeta^\prime}{\zeta^\prime}.\label{psi_h_func}
\end{equation}
Here, $\phi_h(\zeta)$ is thermal stability function that represents dimensionless form of potential temperature gradient, $\phi_h(\zeta)=(\kappa z/T_*)\partial \theta/\partial z$. 
\textcolor{black}{The stability correction functions\cite{Chenge_Brutsaert_2005, Brutsaert_2005} for potential temperature are provided in Appendix A.}

For the potential temperature field, the most common surface thermal boundary conditions used for LES of atmospheric boundary layers are specified surface temperature (i.e. specifying $\theta_s$) or a specified heat flux. However, since turbine wakes are expected to affect locally the heat transfer coefficients in wakes and that the thermal inertia of the ground will play a role, it is  likely that specifying either of these types of boundary conditions will miss important localized features of the flow. Therefore, similarly to the approach taken by Lu and Porté-Agel \cite{lu2015impact}, we solve an additional evolution equation for $\theta_s$, the temperature field in the soil.  A surface thermal energy balance is used for determining land-surface temperature and surface heat flux.  
A similar method was used by Deardorff \cite{deardorff1974three}, who assumed horizontal homogeneity and solved a single one-dimensional heat equation in the soil. Allowing for spatial dependence, Abkar, Sharifi, and Porté-Agel \cite{lu2015impact,abkar2016wake} solved a 3D soil conduction equation. However, given the dominance of vertical heat conduction over horizontal heat conduction in the soil, we only solve a one-dimensional (1D) differential equation for the soil domain. In this approach, the soil heat flux, the flux of sensible heat and an imposed net radiative flux are included in the surface energy balance. Specifically, at every horizontal grid-point of the simulation, we solve the 1-D heat equation for $Z_s < z < 0$,
\be
\frac{\partial \theta_s}{\partial t} = \alpha_s \frac{\partial^2 \theta_s}{\partial z^2},
\label{eq:1dsoilconduction}
\ee
where
$\alpha_s$ is the diffusivity coefficient of the soil. $Z_s$ is the depth of the soil domain. The following boundary conditions are used: at the ground surface at $z=0$, we balance incoming radiation and convective heat flux with heat conduction into the soil:
\begin{equation}
    k_s \frac{\partial \theta_s}{\partial z}\Big|_{z=0}(x,y,t) = H_{\rm  rad}(t) - H_{\rm air}(x,y,t), ~~~~~~~ \rm with ~~\,\,H_{\rm  rad}(t) = \max \left[B_0\sin \left(\frac{2\pi \, t}{T}\right) - B_2,B_1\right].
    \label{eq:heat flux balance}
\end{equation}
\textcolor{black}{Here, $H_{\rm air}=c_{\rm air}\,q_{\rm {3|surf}}$,} while we assume an adiabatic boundary condition far below the ground at $z=Z_s$ (where $Z_s=-1.40$ m): 
\begin{equation}
    \frac{\partial \theta_s}{\partial z} \Big|_{z = Z_s}(x,y,t) = 0.
\end{equation}

\begin{figure}[H]
\centering
\includegraphics[width=\textwidth]{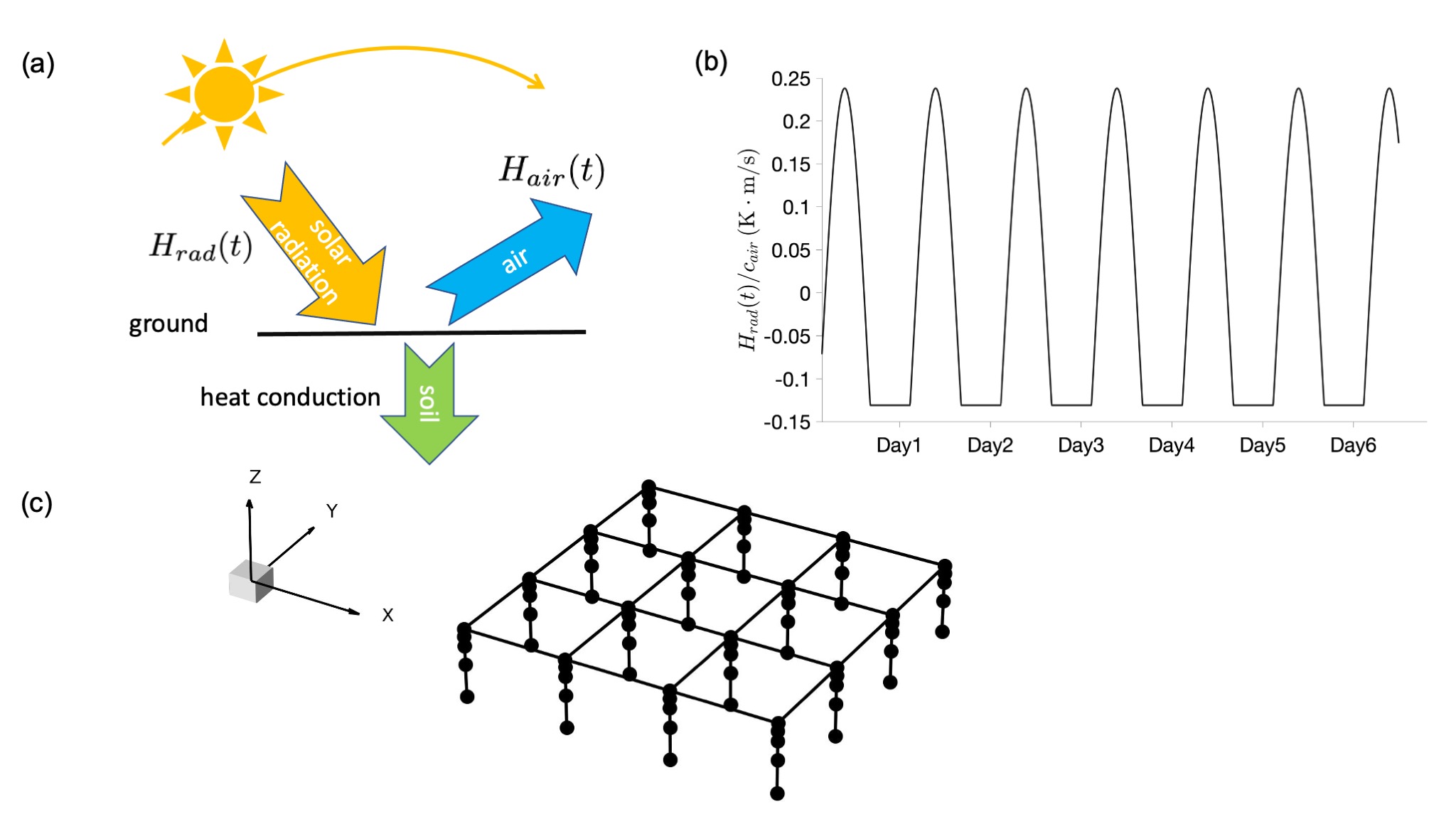}
\caption{Ingredients of the soil conduction model used in LES. (a) Sketch of the thermal energy partition of radiative heat-flux at the ground surface partitioned into thermal conduction in  the soil and thermal convection in  the air. (b) Time series of imposed radiative heat-flux with zero mean. (c) Illustration of the 1D grids used to solve the 1D heat conduction equation at every LES grid point separately. Horizontal conduction is neglected as justified by the thermal penetration depth during a 24 hour cycle being much thinner than the horizontal grid spacing in the LES.}
\label{fig:diurnal_soil_model_sketch}
\end{figure}

The initial soil temperature is uniform and equals to 295 $\text{K}$. \textcolor{black}{The soil is treated as effectively dry 
%to remain consistent with the dry-air boundary. We prescribe its 
with a volumetric heat capacity of $c_s$ = 1.85 $\times10^6\, \text{J}/\text{m}^3\cdot \text{K}$,   thermal diffusivity   $\alpha_s$ = $5.0 \times 10^{-7}\,m^2/s$, and   thermal conductivity   $k_s = c_s\alpha_s = 0.925\, \text{J}/\text{m}\cdot \text{s}\cdot\,\text{K}$ \cite{deardorff1974three}. The soil’s effective ``dry'' parameterization assumes negligible moisture effects.}
%Eq. \ref{eq:heat flux balance} results from the heat flux balance at the ground surface.
%$H_{\rm air}$ is given by Eq. \ref{eq:air turbulent heat flux} and 
$H_{\rm  rad}(t)$ is the imposed representative radiative flux throughout the day, with $B_0 = \textcolor{black}{371.5}\,\text{W}/\text{m}^2$, $B_1 = \textcolor{black}{-157}\,\text{W}/\text{m}^2$, $B_2 = \textcolor{black}{85.5}\,\text{W}/\text{m}^2$, and $T = 24\,\text{hr}$. The imposed radiative flux is \textcolor{black}{guided} by the Cooperative Atmospheric Surface Exchange Study (CASES-99) field campaign \cite{poulos2002cases} \textcolor{black}{and is comparable in magnitude to that} used in the original diurnal cycle LES approach of Kumar {\it et al.} \cite{kumar2006large}. Note that the units of radiative flux can be expressed as either $\text{W}/\text{m}^2$ or $\text{K} \cdot \text{m}/\text{s}$, and they are related through the volumetric heat capacity of air, $c_{\rm air} = \textcolor{black}{1{,}200}\, \text{J}/\text{m}^3 \cdot \text{K}$. However, the time-history of radiative heat flux is shifted such that it has zero mean over a 24-hour period to enable reaching daily time-periodic conditions.
Fig. \ref{fig:diurnal_soil_model_sketch} illustrates the approach.

The thermal penetration depth estimated using $\delta_{\rm s} \sim \sqrt{\alpha_s/\omega}$ with $\omega = 2\pi/24\,\text{hr}^{-1}$ is on the order of $\delta_{\rm s} \sim 0.08\,\text{m}$. This depth is much smaller than the horizontal resolution in the LES, thus justifying the 1D approximation to integrate Eq. \ref{eq:1dsoilconduction} as well as placing the lower boundary condition at a much larger depth of -1.4m. Numerically, ${\partial^2 \theta_s}/{\partial z^2}$ is discretized using a stretched-grid finite difference scheme, with the second derivative at $z = z(k)$ approximated as: 
\be\label{eq:soil_gradient}
\frac{\partial^2 \theta_s}{\partial z^2} |_{z = z(k)} = \frac{2}{[z(k+1) - z(k-1)]} \left[ \frac{\theta_s(k+1) - \theta_s(k)}{z(k+1) - z(k)} - \frac{\theta_s(k) - \theta_s(k-1)}{z(k) - z(k-1)} \right],
\ee
where $k$ represents the $k$-th grid point. The vertical grid discretizations into 31 levels of soil temperature, extending to a depth of $Z_s \approx -1.4\,\text{m}$, uses a stretched configuration, $z(k) = z(k-1) + \Delta_{\rm surface} \cdot \sigma^{k-1}$, with a surface grid spacing $\Delta_{\rm surface}$ of $0.001\,\text{m}$ and a stretch ratio $\sigma$ of 1.2, resulting in finer spacing near the surface. The heat equation is solved along $z<0$ for $\theta_s(k,t)$ using the Adams-Bashforth 2nd-order scheme for time advancement at each time step. Note that Eq. \ref{eq:soil_gradient} was implemented using simple explicit formulation as the time step was much shorter than any relevant diffusion timescale. \textcolor{black}{ Extensive validation tests (e.g. offline calculations using much finer grids and time-steps) confirm that the resolution employed yields highly accurate solutions of the 1D heat equation in the soil}.
%\textcolor{black}{Validation tests of the soil model solver are presented in Appendix B.}

\subsection{Wind turbine model\label{sec:wind turbine}}

During an initialization stage of the simulation using a coarse grid (to be described in the next section), we use an actuator disk representation with a local thrust coefficient of $C_T^\prime = 1.33$.  \textcolor{black}{$C_T^\prime$ is defined based on the disk-averaged velocity $u_d$ at the rotor location\cite{calaf2010large}, according to $F_d = \frac{1}{2} C_T^\prime u_d^2 A_d$ (where $F_d$ is the force per unit mass acting on a disk area $A_d$). Unlike the far-upstream velocity $U_{\infty}$ usually used to define the thrust coefficient $C_T$, the local velocity $u_d$ is immediately available in LES. }
For the production run and data generation during the target 24 hour period of the diurnal cycle, we use a finer grid resolution, and an advanced actuator line model (ALM)   \citep{sorensen2002numerical,martinez2015large,martinez2024generalized} is used. In the ALM method, each rotating blade is represented by a line of points (here we use $N=100$ ALM points) at which the ALM force per unit length is computed as  
\begin{equation}
    \mathbf{f}_{\rm alm} = 0.5   c |\mathbf{V}_{\rm rel}|^2(C_L \mathbf{e}_L + C_D \mathbf{e}_D),
\end{equation}
where  $|\mathbf{V}_{\rm rel}|$ is the magnitude of the relative velocity of the upwind flow to the wind turbine blade, $c$ is the local airfoil chord length, $C_L$ and $C_D$ are lift and drag coefficients obtained from tabulated airfoil data, respectively, while $\mathbf{e}_L$ and $\mathbf{e}_D$ are unit vectors along the direction of the lift and drag forces at each actuator point, respectively. These forces are then smeared using a Gaussian kernel to project them into the computational LES grid using a width $\epsilon = 2 (\Delta_x \Delta_y \Delta_z)^{1/3}$ as recommended in the literature\citep{troldborg2009actuator,martinez2015large}. 
Since $\epsilon$ is substantially larger than the optimal $\epsilon$ required to capture the local aerodynamic interactions causing e.g. tip losses, we use a correction developed from the generalized filtered lifting line theory \citep{martinez2019filtered, martinez2024generalized}. This model enables is us to combine ALM with a relatively coarse LES resolution. 

The turbine parameters used in this study are based on the reference NREL-5MW baseline wind turbine \cite{jonkman2009definition}.  The three blade turbine has a diameter of $D=126\,\text{m}$ and its hub height is at elevation $z_h = 90\,\text{m}$. Its standard tabulated lift and drag coefficients are used.  
\textcolor{black}{The ALM implementation in LESGO allows to model detailed turbine operation controls, such as feathering (pitching) the blades during region III operations, e.g. above rated conditions. However, in the current simulation the turbines are operated during the entire time at optimal tip-speed ratio (``region II'' operation, also without including regions 1.5 and 2.5), in order to avoid having to store additional data relating to curtailment (e.g. blade pitch) and possibly complicated turbine control actions. We therefore refer to the modeled turbine as the NREL-5MW+ turbine, meaning that it is allowed occasionally to rotate slightly faster than the maximum rotation rate of the original NREL-5MW reference turbine (up to about 25-30\% faster, during short periods of time during the day)}. More details of the turbine and ALM model implementation are provided in Zhu {\it et al.} \cite{Zhu2025jhtdbwindpaper}.

\section{Simulation setup and data acquisition}
\label{sec:simulationdata}
\subsection{Simulation setup}

The computational domain and dimensions are illustrated in Fig. \ref{fig:schematic1}.  
The wind farm configuration consists of two columns and four rows of \textcolor{black}{NREL-5MW+}   turbines. The turbine spacings in the streamwise ($x$) and spanwise ($y$) directions are $7D$ and $5D$, respectively, where $D = 126\,\text{m}$ denotes the rotor diameter. The domain width, $L_y$, is chosen as $10D = 1{,}260\,\text{m}$, with the centers of the two turbine columns positioned $2.5D$ from the nearest side computational domain boundaries, ensuring that turbine center locations are positioned at grid points. In the streamwise direction, the wind farm domain extends over a length of $L_x = 63D = 7{,}936\,\text{m}$. This includes a length of $14D$ upstream of the first turbine row and $20D$ downstream of the last row to provide sufficient space for capturing the wind farm blockage effect upstream and allowing the wake flow to fully develop downstream. Additionally, a fringe region of $8D$ is included at the downstream end to prevent any outflow from influencing the inflow due to the horizontally imposed periodic boundary condition. The precursor domain uses exactly the same domain size and resolution as the entire wind farm domain. 

The flow is forced via a prescribed constant geostrophic wind velocity with magnitude $G=(U_g^2+V_g^2)^{1/2}= 15\,\text{m}/\text{s}$ and angle $\alpha_g=\arctan(V_g/U_g)= -28^{\circ}$ with respect to the $x$ direction. This angle is chosen so that the average wind direction over the 24 hr period at turbine hub height is (approximately) in the $x$ direction. The (mid-latitude) Coriolis parameter is $f_c = 1\times 10^{-4}\, \text{s}^{-1}$. The aerodynamic ground surface roughness is set to $z_0 = 0.1\,\text{m}$, and the reference potential temperature is specified to be $\theta_0 = 295\,\text{K}$. The simulation is initialized with a constant streamwise velocity $U_g = 15\,\text{m}/\text{s}$ and zero velocity for the spanwise and vertical components. The potential temperature is initialized with a uniform temperature of $295\,\text{K}$ in the lowest $1000\,\text{m}$. Above this layer, the ABL is set to be laminar with a constant lapse rate of $10\,\text{K}/\text{km}$.  In the lower layer below $1000\,\text{m}$, small random (white noise) perturbations are added to the initial velocity and potential temperature fields. The initial soil temperature is uniform and equals $295\,\text{K}$.  

Simulations are initially conducted over approximately five full diurnal cycles using a coarse resolution to ensure convergence and establish periodicity from one day to the next. Subsequently, an additional diurnal cycle is simulated with a finer spatial resolution. The table \ref{tab:tabledomain} displays relevant information regarding the computational domains of the precursor and the wind farm regions.  The database construction and subsequent data analysis focuses on this last diurnal cycle obtained from the fine resolution simulation.  During the initial stages using the coarse resolution, the wind turbines are represented using the actuator disk model (ADM) and for simplicity of implementation, the disks are not yawed during the several days of evolution simulated on the coarse grid. 
\textcolor{black}{Then the horizontal grid spacing is halved, the actuator line model (ALM) is activated, and the turbines are allowed to yaw to align with the mean hub-height incoming velocity from the precursor domain. After switching from coarse to fine grid, and from ADM to ALM, the simulation is run for 4 hours to build up small scales and flush out any remnants of the ADM turbine representation (4 hours is on the order of 16 flow-through times). Then, at what corresponds to a time of 15:00 hr, the data storage is started.} 

Once the actuator line model  is activated in the fine-resolution simulations, for each turbine  the angular velocity $\Omega$ is set such that the tip speed ratio $\lambda$ remains fixed. \textcolor{black}{As explained before, we purposefully  operate the  NREL-5MW+}   turbines in ``region II'' throughout the diurnal cycle to avoid additional controller actions such as curtailment or cut-off conditions that would add additional complexity to characterization of turbine operation. The tip-speed ratio is fixed near the optimal operating point of the turbine, namely $\lambda= 7.5$. The tip-speed ratio is calculated using the upstream axial velocity $U_{\rm ax}$, obtained from LES results evaluated over each wind turbine swept region, but shifted one grid cell upstream.  Finally then, the angular velocity is given by $\Omega = 7.5 \times R/ U_{\rm ax}$.

\begin{table}[H]
\centering
\caption{Computational domain parameters}
\begin{tabular}{|c|c|c|c|}
\hline
Cases & \parbox[t]{5.5cm}{\centering Precursor + wind farm domain size\\ $L_x \times L_y \times L_z$ (m)} & \parbox[t]{3.5cm}{\centering Number of grid points  \\ $N_x\times N_y\times N_z$} & \parbox[t]{3cm}{\centering Grid resolution  \\ $\Delta x \times \Delta y\times \Delta z$ (m)} \\
\hline
coarse & $15{,}876\times 1{,}260\times 2{,}400$ & $432\times 72\times 240$ & $36.75\times 17.5\times 10$   \\ \hline
fine & $15{,}876\times 1{,}260\times 2{,}400$ & $864\times 144\times 480$ & $18.375\times 8.75\times 5$   \\ \hline
Sampled data & $12{,}899.25\times 1{,}260\times 2{,}400$ & $351\times 72\times 480$ & $36.75\times 17.5\times 5$   \\ \hline
\end{tabular}
\label{tab:tabledomain}
\end{table}

\begin{figure}[H]
\centering
\includegraphics[width=0.8\textwidth]{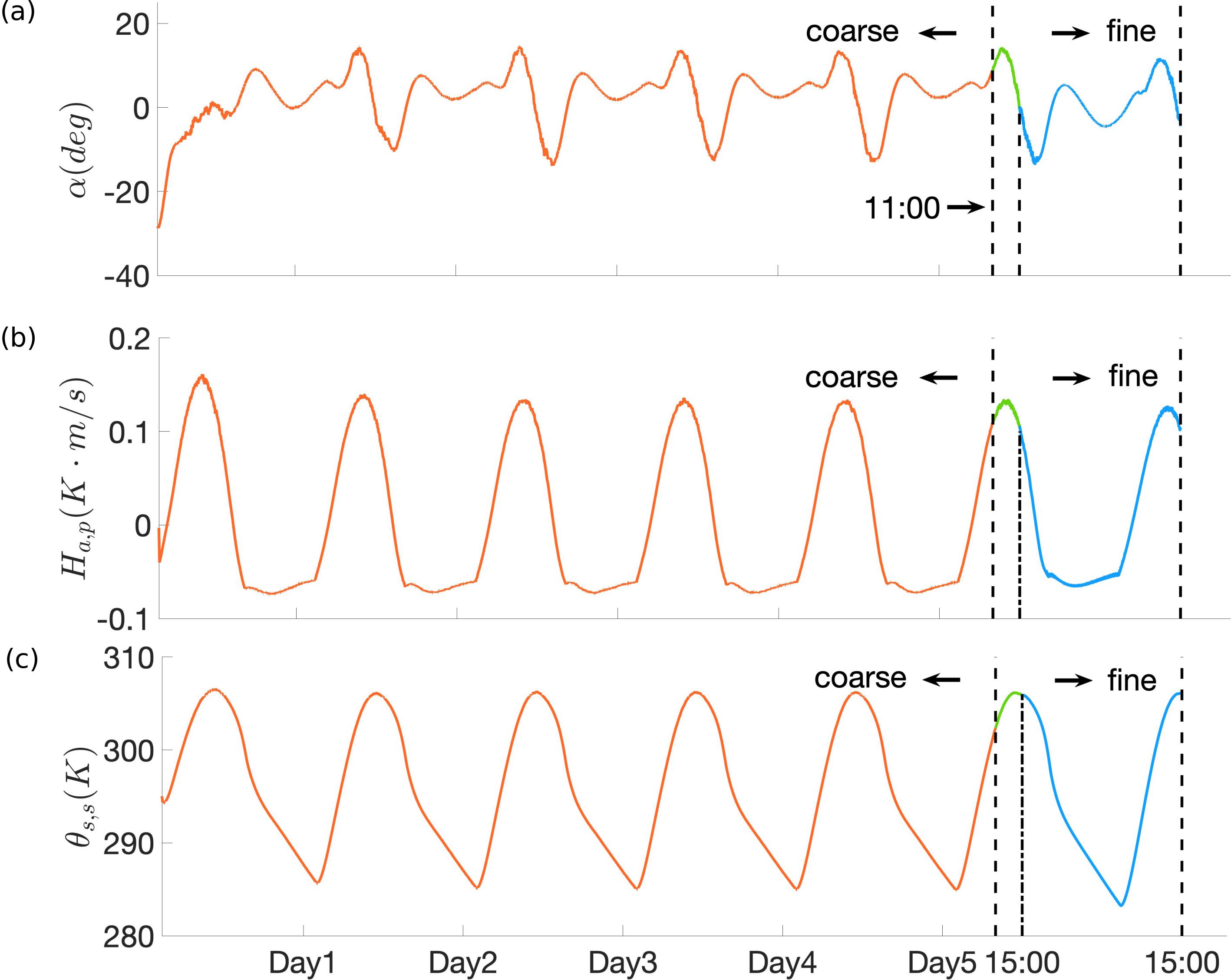}
\caption{Time series of three quantities from the coarse-resolution simulation in the precursor domain during the initial five diurnal cycles, exhibiting convergence towards 24-hour periodic behavior. (a) Horizontally averaged wind direction relative to the streamwise direction in the precursor domain. The  orange  line shows the results from the coarse resolution, \textcolor{black}{the green line shows the results after switching to fine resolution and ADM to ALM wind turbine modeling}, while the solid blue line shows the result for the last 24 hours using fine resolution, \textcolor{black}{starting at 15:00 of day 5 when data collection takes place}. (b) Horizontally averaged heat flux into the air in the precursor domain from the precursor and coarse resolution run. (c) Horizontally averaged ground surface temperature in the precursor domain during the coarse resolution run.}
\label{fig:diurnal_coarse}
\end{figure}
Based on the coarse-grid initial simulations, Fig. \ref{fig:diurnal_coarse}(a) shows time histories of the spatially averaged flow angle at hub-height $z_h = 90\,\text{m}$, computed at every time-step in the precursor domain according to $\alpha = \arctan(\langle v\rangle_{x,y}/\langle u\rangle_{x,y})$. Averaging is done   over horizontal $x,y$ planes in the precursor domain. The last 24 hours correspond to the fine resolution used for data analysis and database. At about 15:00 hr of the last period at the start of the fine resolution data acquisition, the flow angle at hub height is nearly zero. Starting the data acquisition for the fine-resolution simulation at a time when the yaw angle is zero minimizes the time required for the wake flow to develop and adjust behind the turbines since the coarse-grid ADM used non-yawing turbines so that the ADM wakes are more realistic when $\alpha\approx 0$. Another observation is that the fine-resolution flow angle displays some differences with the values of the preceding cycle at coarse resolution, especially near midnight where the decrease in angle is more pronounced. Also the soil temperature reaches a lower minimum value (at early morning hours) in the fine grid simulation as compared to the coarse ones. These observed differences are due to the grid resolution that affects   the behavior especially during the stably stratified night-time conditions in which turbulent flow structures occur at much smaller scales. Still, at the end of the 24 hour period of the fine grid simulation, the end values return to their initial values at \textcolor{black}{15:00 hr} of the preceding day. We conclude that the fine resolution run can be considered to have converged sufficiently to periodic behavior.  

Also shown in Fig. \ref{fig:diurnal_coarse} (b) and (c) are the   surface heat flux and surface soil temperature averaged over $x,y$ plane in the precursor domain. As can be seen by around 120 hours (about 4 days) a time-periodic flow condition has been fully achieved.  At 11:00 hr (approximately 125 hours after the start of the coarse-grid simulation) of the coarse simulation, fields are interpolated (spectrally with zero-padding in the spectral directions and using linear interpolation in the $z$ direction) onto the finer mesh, and the ALM model is activated to replace the ADM. The fine-grid simulation is run for 4 hours, corresponding to around 10 flow-through times across both the precursor and wind farm domains, reaching an equilibrated state on the fine grid. At 15:00 hr, the simulation continues in order to generate a 24-hour run.  The simulation time-step is held constant for the purpose of storing the data, and is set to $\Delta t_{\rm les}=0.05\,s$. This time-step is small enough to conform to the CFL condition during the entire simulation time.

\subsection{Data acquisition for database}
\begin{figure}[H]
    \centering
    \includegraphics[width=\textwidth]{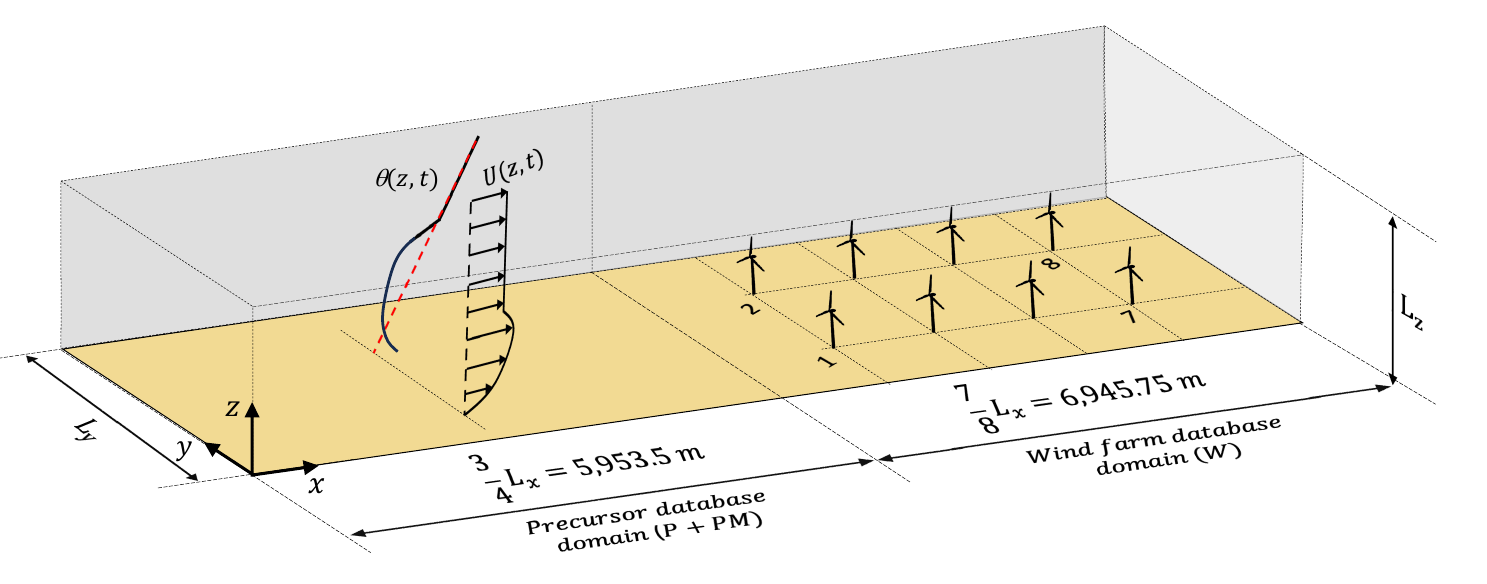}
    \caption{Schematic representation of the database domain. It represents the physical domain available in the database, merging the precursor domain ($P+P_M$) up to the end of the mapping region at $3/4 L_x$, with the windfarm domain ($W$) but excluding the fringe region. Turbines are numbered from 1 to 8 as shown. The domain dimensions are $1.625 \, L_x$ (streamwise) = $3L_{x}/4$ (precursor) + $7L_{x}/8$ (wind farm), $L_y$ (spanwise), and $L_z$ (wall-normal).
    }
    \label{fig:schematic2}
\end{figure}

To create the database for subsequent analysis, field variables are spatially filtered at scale ($2 \Delta x$, $2 \Delta y$) and coarse-grained in horizontal planes, thus reducing the amount of data that needs to be stored by a factor of 4. This approach is justified on the grounds that the finest resolved scales in LES are those most affected by numerical and modeling uncertainties.  For the stored dataset, the precursor and wind farm domains are combined to create seamless fields. The three velocity components, potential temperature, pressure, and turbulent viscosity are merged into a unique domain containing both the periodic inflow as well as the wind farm domain. The fringe regions at the outlet of both the precursor (with a length of $L_x/4$)  and wind farm domains (with a length of $L_x/8$) are not stored, as the field data there are not physical there.  The resulting 3D database \textcolor{black}{domain} is illustrated in Fig. \ref{fig:schematic2}. Due to the non-local nature of the Poisson equation used to solve for pressure and properties of the concurrent precursor simulation method in which only the velocities but not the pressure fields are coupled, the stored pressure field contains a small offset discontinuity at the merger plane between the two domains. 

Data are stored at every 10 time steps, i.e. $\Delta \,t_{\rm db} = 10 \Delta \, t_{\rm les}=0.5\,s$. The ALM turbine data at each of the ALM actuator point along the three blades of each of the 8 turbines are stored at the original LES time-step $\Delta \, t_{\rm les}=0.05\,s$.  Additional details regarding which flow and wind turbine variables are stored in the database are provided in Appendix B. 

\section{Diurnal cycle: general observations}
\label{sec:generalobs}

\begin{figure}[H]
\centering
\includegraphics[width=\textwidth]{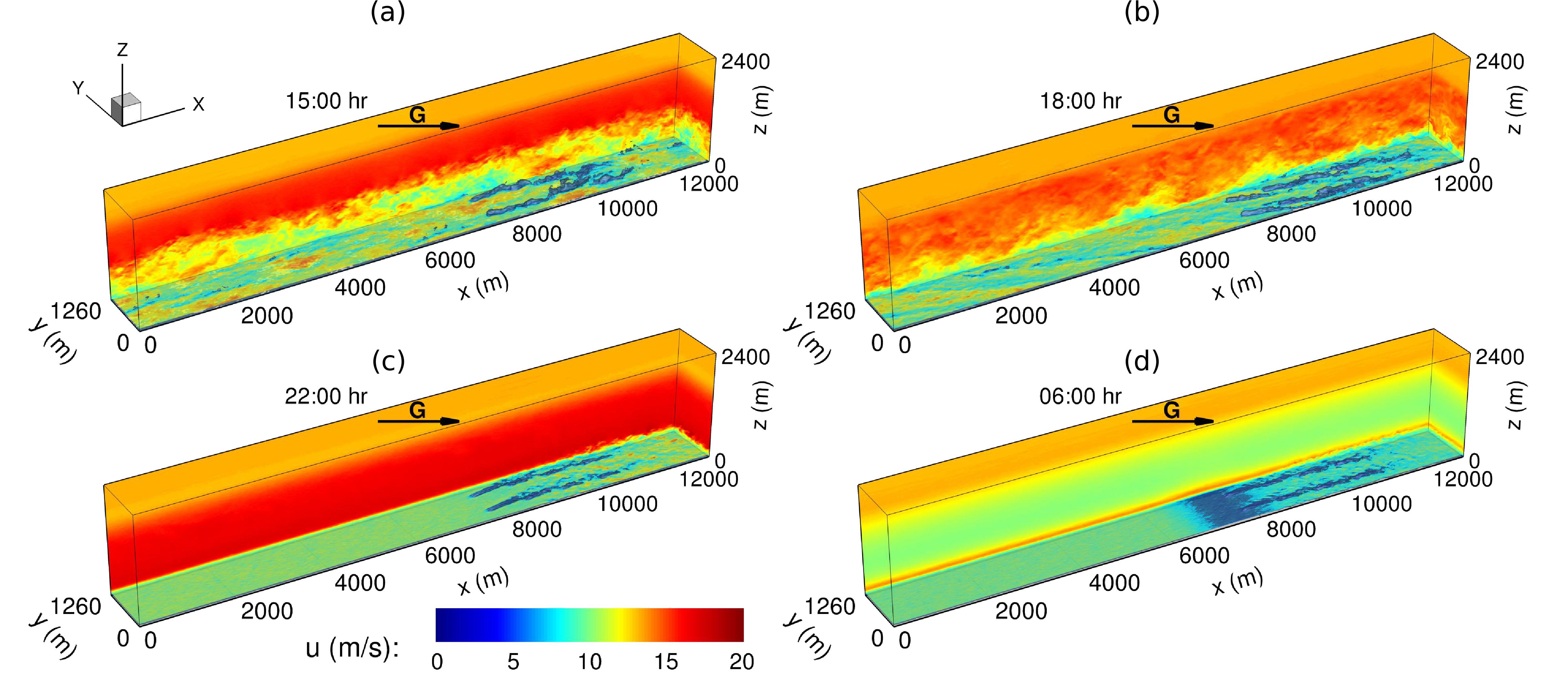}
\caption{Snapshots of x-direction ($u$) velocity snapshots at four times during the diurnal cycle, combining fields in both the precursor and wind farm domains. (a) Mid afternoon (15:00 hr). (b) Late afternoon (18:00 hr). (c) Evening (22:00 hr). (d) Early morning (06:00 hr). The boundary between the precursor and wind farm  domains is located at $x=5{,}953.5$ m, with seamless transition of velocity fields between the two domains. The bottom plane shown is at height $z = 70\,\text{m}$. The arrow at the top indicates the imposed geostrophic wind $G$, oriented in such a way that the daily average flow direction at hub height is in the $x$ direction.}
\label{fig:diurnal_layout}
\end{figure}

Overall, as will be seen, the trends discussed in this section are qualitatively in good agreement with results and observations from prior studies of the ABL during a diurnal cycle \cite{kumar2006large,kumar2010impact,quon2024measurement,abkar2016wake,sharma2017perturbations}.  Fig. \ref{fig:diurnal_layout} shows four snapshots of the $x$ direction velocity component from LES on several planes in the flow. The figure shows that, over the diurnal cycle, both the flow direction and ABL height change substantially. At a time of 18:00 hr, the flow direction near the ground is about $\alpha = - 15^{\circ}$, consistent with the observed orientation of the bottom streaks. Also visible are the turbine wakes and the sharp inversion layer near $z_i \approx 1{,}400$ m at that time near the end of the day. Conversely, in the morning time at 06:00 hr,   significantly reduced velocity develops in front of the wind farm due to the wind farm blockage effect, and \textcolor{black}{the nocturnal low-level jet can be observed   at low elevations (around $z\sim 200$m), characterized by a narrow ($\Delta z\sim 150$m) band} of high-speed wind. Both phenomena will be discussed further in section \ref{sec:resultspower}.
 
\begin{figure}[H]
\centering
\includegraphics[width=\textwidth]{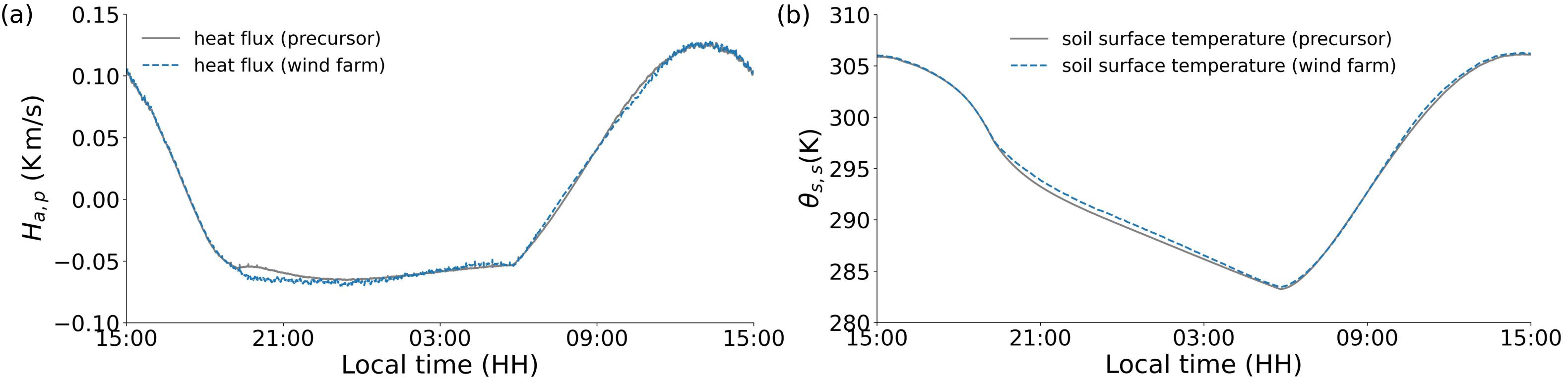}
\caption{Time series of thermal properties at the land-atmosphere interface (bottom surface) in the precursor \textcolor{black}{and wind farm domains}, over the diurnal cycle. (a) Horizontally averaged heat flux into the air. (b) Horizontally averaged ground surface temperature.}
\label{fig:diurnal_thermal}
\end{figure}

 Fig. \ref{fig:diurnal_thermal} shows time series of surface heat flux (to air) and soil surface temperature over the 24 hour period in the precursor domain. As can be seen, start and ending values are the same, consistent with temporal periodicity. The average heat flux is nearly zero, also consistent with periodic behavior.  
 \textcolor{black}{Fig. \ref{fig:diurnal_thermal} also shows results in the wind farm domain, evaluated between the first turbine row and a location \(7D\) downstream of the last row. As can be seen, some differences exist between the mean values in the precursor and wind farm domains, especially for the heat flux at around 20:00 hr where a difference of about 20\% is observed. While the farm primarily redistributes surface flux and skin temperature locally,  mainly affecting the spatial pattern of near-surface thermal variables, even the average over the domain is affected. In the absence of a soil model, if we prescribed a homogeneous surface heat flux equal to that of the radiative flux, it would be exactly the same in both the precursor and wind farm domains.}
 
\begin{figure}[H]
\centering
\includegraphics[width=\textwidth]{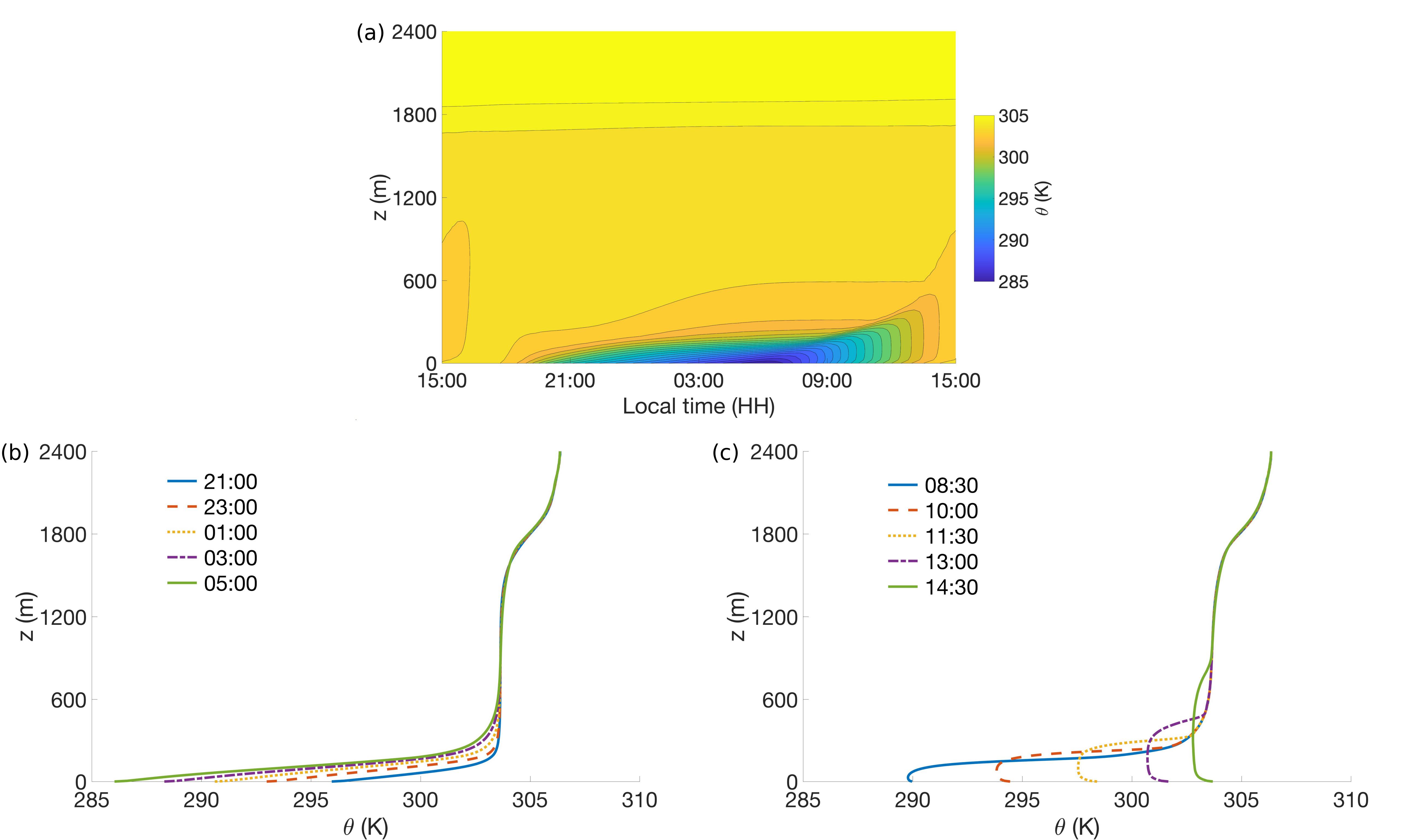}
\caption{Spatio-temporal evolution of horizontally averaged potential temperature in the precursor domain during the diurnal cycle, using a contour plot (a) and profiles at various times from late evening to early morning (b) and from early morning to early afternoon (c).}
\label{fig:diurnal_potential_temperature}
\end{figure}

 Figure \ref{fig:diurnal_potential_temperature} shows the spatio-temporal evolution of the mean potential temperature (averaged in horizontal directions) across the entire boundary layer. After approximately 18:00 hr, when the net radiative flux becomes negative, as shown in Fig. \ref{fig:diurnal_thermal}a, a gradual temporal decrease in potential temperature and stable thermal stratification above the ground surface are observed, which is also clearly exhibited in the individual profiles at various times shown Fig. \ref{fig:diurnal_potential_temperature}b. During the day, solar radiation heats the ground surface and the air above it, leading to the growth of the boundary layer, as explicitly shown in Fig. \ref{fig:diurnal_potential_temperature}c.

\begin{figure}[H]
\centering
\includegraphics[width=\textwidth]{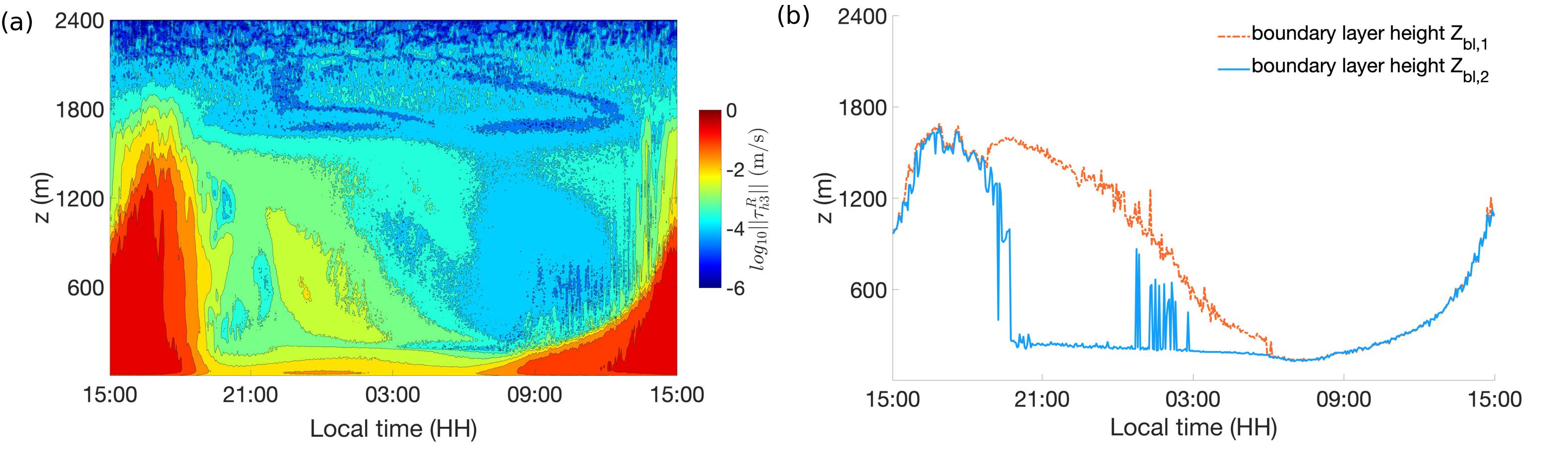}
\caption{(a) Contour plot of the horizontally averaged  Reynolds stress distributions from the fine-resolution simulation in the precursor domain and (b) the derived ABL height over a diurnal cycle.  The ABL height $Z_{bl}$ is determined as the location where the horizontally averaged vertical Reynolds stress is $10\%$ of its instantaneous maximum value. If multiple points meet this criterion at a given snapshot, the highest one is selected for $Z_{bl,1}$, and lowest one for $Z_{bl,2}$.}
\label{fig:diurnal_stress_ABL}
\end{figure}

In order to quantify the boundary layer height we examine spatio-temporal contours of \textcolor{black}{resolved} Reynolds shear stress distribution across the diurnal cycle. The Reynolds shear stress  magnitude is defined as follows

\be
|| \tau^{R}_{h3}||(z) = \left( \langle u'w'\rangle^2 + 
\langle u'v'\rangle^2 \right)^{1/2}.
\ee
\textcolor{black}{The covariances $\langle u'w'\rangle$ and 
$\langle u'v'\rangle$ are computed at each saved time $t$ and height $z$ using horizontal (plane) averaging over the $x,y$ planes. Note that we do not include the subgrid-scale component of the shear stress as it is much smaller than the resolved part.} 
The result is shown in Fig. \ref{fig:diurnal_stress_ABL}a \textcolor{black}{using logarithmic scale}, and conforms to standard \textcolor{black}{knowledge} regarding the diurnal evolution of a standard atmospheric boundary layer. Around sundown at \textcolor{black}{19:00 hr} the boundary layer turbulence rapidly collapses and a stably stratified boundary layer develops near the ground, with heights of between 150-200 m, i.e. of similar height as the wind turbines.  Measures of boundary layer height are known to be difficult to define based on such data. In Fig. \ref{fig:diurnal_stress_ABL}b we show two different definitions that appear consistent with the qualitative features of the contour plot in  Fig. \ref{fig:diurnal_stress_ABL}a: Both the first and the second value for $Z_{bl}(t)$ is given by  $|| \tau^{R}_{h3}||(z=Z_{bl}(t)) = 0.1  {\rm max}_z|| \tau^{R}_{h3}||(z,t)$, where the max is taken separately at each time $t$. There might be multiple points that meet the above criteria at each time, and the largest height is taken for $Z_{bl, 1}(t)$ and the smallest height is taken for $Z_{bl, 2}(t)$. Between these two heights we have the residual layer with reduced mixing and low turbulence but velocities and temperature fields somewhat decoupled from the bottom layer. 

\begin{figure}[H]
\centering
\includegraphics[width=\textwidth]{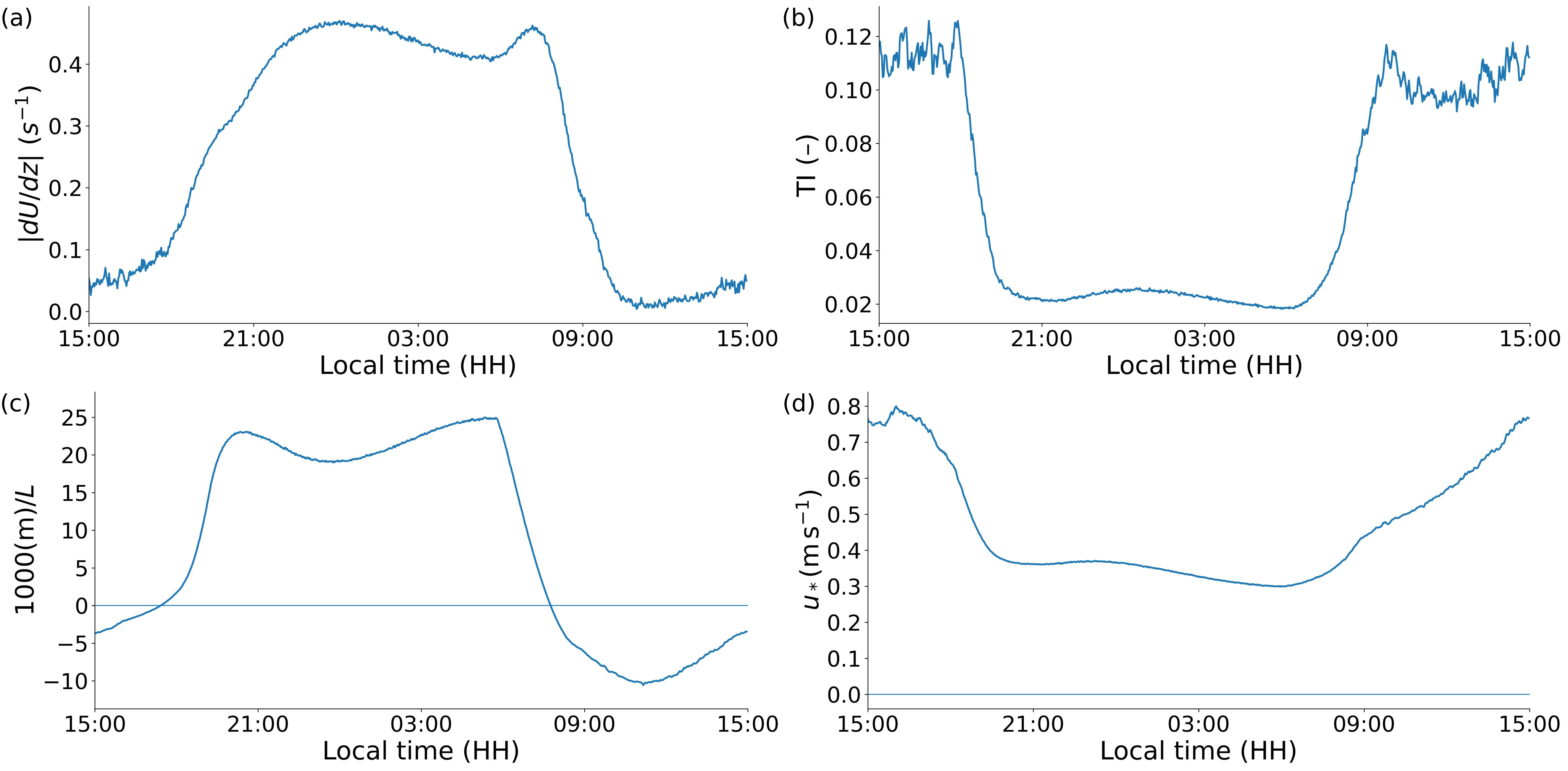}
\caption{Diurnal evolution of boundary-layer diagnostics in the precursor domain. (a) Vertical shear magnitude $|dU/dz|$ at hub height; (b) turbulence intensity $\mathrm{TI}$ at hub height; (c) stability parameter $1000\,(\mathrm{m})/L$ based on the Monin--Obukhov length $L$ (horizontal zero line marks neutral; positive stable, negative unstable).  $1000\,\mathrm{m}$ is chosen because it is the boundary-layer height at 15{:}00 hr as shown in Fig. \ref{fig:diurnal_stress_ABL}b; and (d) friction velocity $u_*$. Curves are horizontal averages over the precursor domain plotted versus local time.}
\label{fig:diurnal_stability}
\end{figure}

\textcolor{black}{Fig.~\ref{fig:diurnal_stability} summarizes the diurnal cycle of boundary-layer diagnostics in the precursor domain. Panel (a) and (b) show the mean velocity shear and turbulence intensity evaluated at hub-height $z=90$m. Figures (c) and (d) show the evolution of inverse Obukhov length and friction velocity (both computed from the heat and momentum flux boundary conditions at the surface) during the diurnal cycle. For plotting purposes, the inverse Obukhov length is non-dimensionalized using a constant reference ABL height of 1000 m.  At night the layer is stably stratified ($1000(\mathrm{m})/L \approx 20\mbox{--}25$); vertical shear at hub height is strong, turbulence intensity at hub height is very small $((\mathrm{TI}\approx 0.02\mbox{--}0.03)$, and friction velocity is slightly lower  ($u_* \approx 0.3\mbox{--}0.4~\mathrm{m\,s^{-1}}$) than during the day. 
Around 08{:}00 hr in the morning the sign of $1000(\mathrm{m})/L$ reverses, indicating the onset of daytime convective mixing, accompanied by a rapid weakening of vertical shear, an increase in $\mathrm{TI}$, and a rise in $u_*$.}

\begin{figure}[H]
\centering
\includegraphics[width=\textwidth]{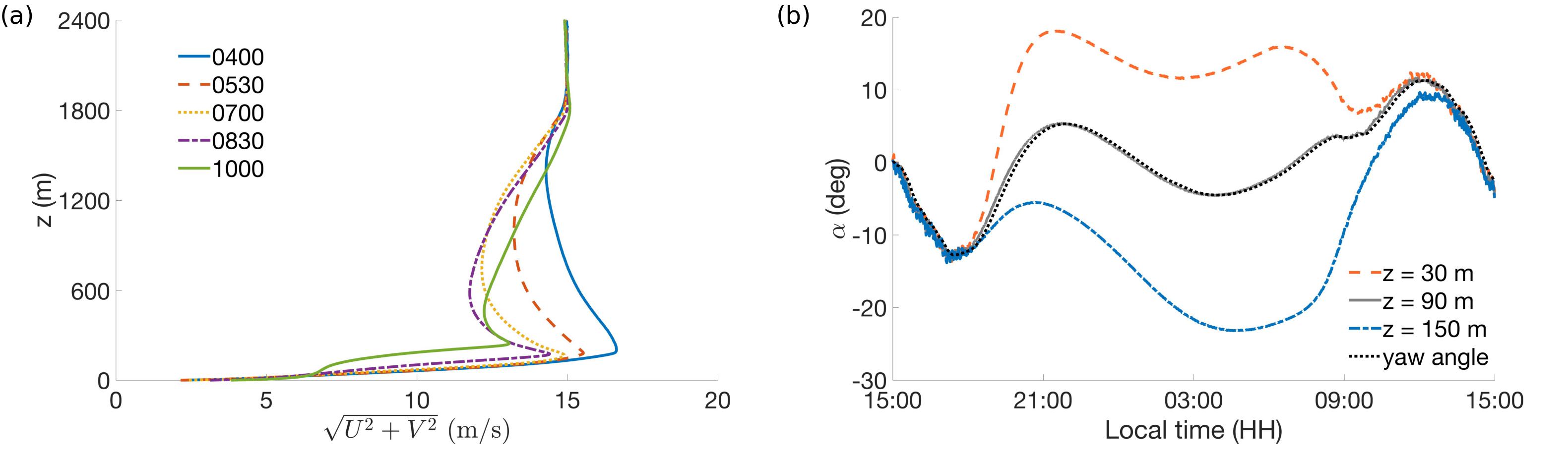}
\caption{Vertical profiles of horizontally averaged velocity from the fine-resolution simulation in the precursor domain and the temporal evolution of the imposed yaw angle for wind turbines over the diurnal cycle. (a) Vertical profiles and temporal evolution of the horizontally averaged velocity at 5 times from 04:00 hr to 10:00 hr, with a \textcolor{black}{a nocturnal low-level jet observed near the ground, at $z \sim 200$m}. (b) Temporal evolution of the horizontally averaged velocity direction relative to the streamwise direction  at three representative heights: $z = 30\,\text{m}$, $z = 90\,\text{m}$ (hub-height), and $z = 150\,\text{m}$, along with the temporal evolution of the imposed yaw angle. The imposed yaw angle for wind turbines is specified based on the time-filtered (20 minutes), horizontally averaged velocity direction at the hub height $z_h = 90\,\text{m}$.}
\label{fig:diurnal_velocity_profiles}
\end{figure}

Fig. \ref{fig:diurnal_velocity_profiles}a shows the evolution of the horizontal velocity magnitude in the early morning. As can be seen \textcolor{black}{a nocturnal low-level jet exists} during the early morning due to nighttime surface cooling, which leads to the development of a temperature inversion and the creation of a stable boundary layer that decouples surface winds from those in the free atmosphere above. 
%Decoupled from surface friction, the winds can move faster resulting in the formation of the jet. 

Additionally, due to variations in boundary layer height throughout the day and the Coriolis effect, the horizontal mean wind direction shifts gradually with altitude, a phenomenon known as wind veer. As shown in Fig. \ref{fig:diurnal_velocity_profiles}b, during the day, as the boundary layer expands and becomes more turbulent, wind speeds and directions become more uniform with height, while at night, the development of a stable boundary layer enhances wind shear and veer. To simplify the yaw angle control of the wind farm, we set all wind turbines to align with the mean wind direction at hub height from the precursor domain (inflow). To avoid unphysical temporal variability,   the precursor domain mean velocity components are filtered using an exponential time-filter with a 20-minute averaging (relaxation) time scale. Hence, only slow veering of the velocity affects the imposed turbine yaw.

\begin{figure}[H]
\centering
\includegraphics[width=\textwidth]{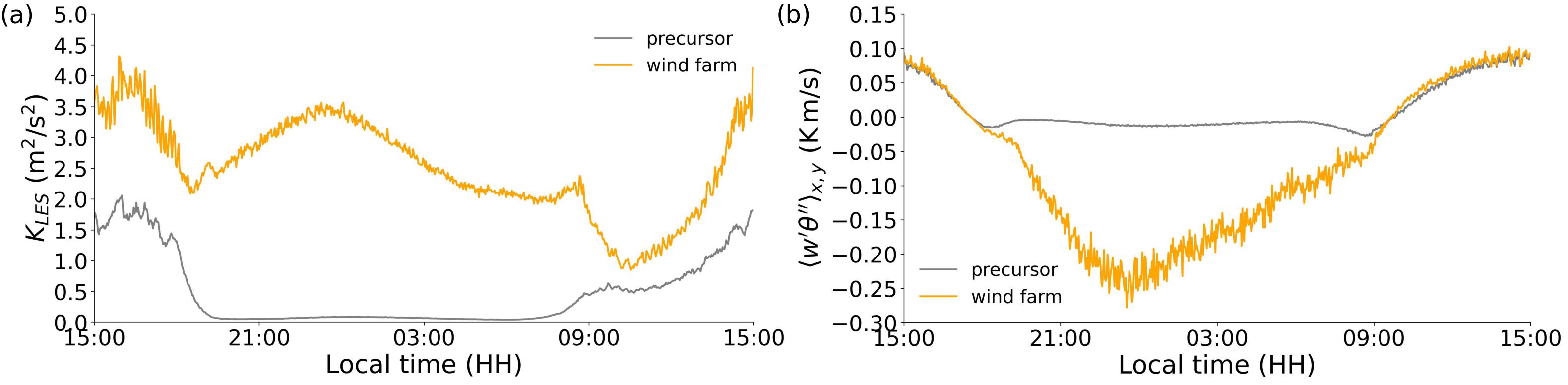}
\caption{Time evolution of the horizontally averaged turbulent kinetic energy (a) and turbulent vertical heat flux  (b) at hub-height from the fine-resolution simulation in the precursor \textcolor{black}{and wind farm} domain over a diurnal cycle.}
\label{fig:diurnal_turbulence_profiles}
\end{figure}

Fig. \ref{fig:diurnal_turbulence_profiles} shows the temporal evolution of horizontal turbulent kinetic energy and vertical turbulent heat flux at  hub height. The   turbulent kinetic energy of the resolved scales in LES, defined as $k_{\rm les} = \frac{1}{2} [\langle u'^2\rangle_{x,y}+\langle v'^2\rangle_{x,y}+\langle w'^2\rangle_{x,y}]$, reaches its maximum value during the day due to solar heating \textcolor{black}{ and buoyancy production}, while at night, it decreases to around 0.1 $(m^2/s^2)$, a small but non-negligible value that still influences wind turbine wakes and residual turbulence. The vertical turbulent heat flux at hub height is positive during the day and becomes negative at nighttime, indicating a net downward turbulent heat flux. In the context of stable stratification at night, this downward flux transports warmer air from higher elevations to lower elevations, thereby warming the air near the ground. However, the vertical turbulent heat flux at hub height in the precursor domain has a magnitude at least one order of magnitude smaller compared to that in the wind farm region, as shown in \textcolor{black}{Fig. \ref{fig:diurnal_turbulence_profiles} and Fig. \ref{fig:dirunal_inst_heatflux}}. Therefore, the warming effect at nighttime is negligible except directly behind the wind turbines, which will be described in detail later.

\begin{figure}[H]
\centering
\includegraphics[width=0.9\textwidth]{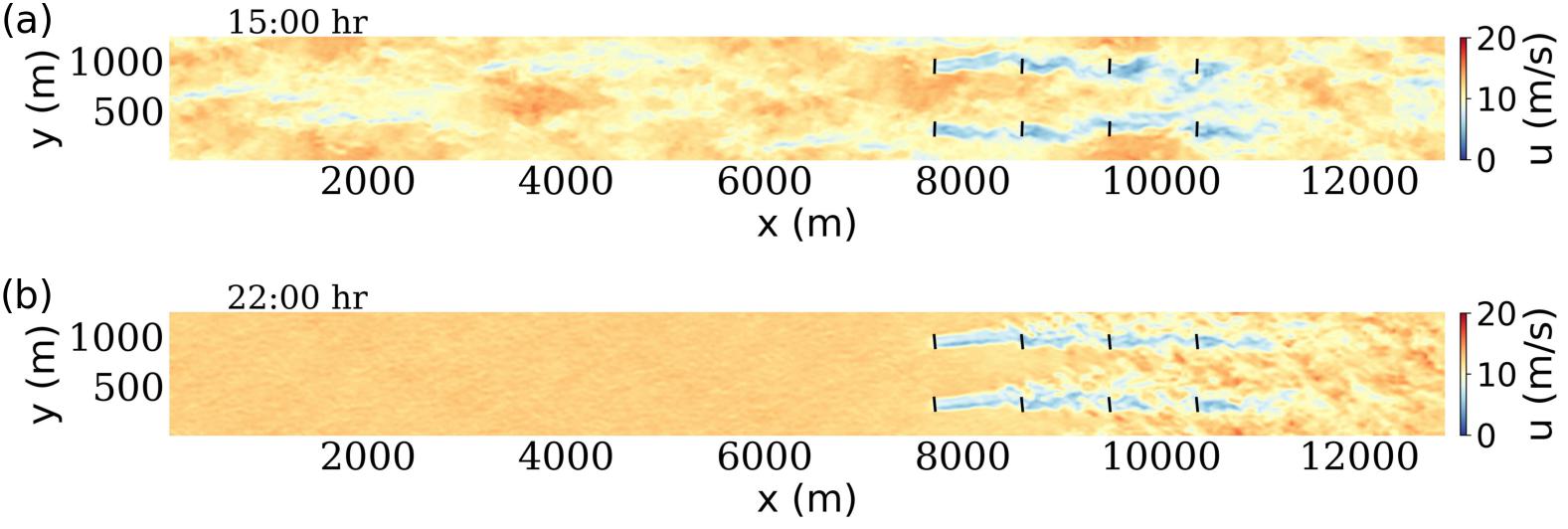}
\caption{Instantaneous streamwise velocity at hub height recorded at (a) afternoon 15:00 hr,   (b) evening 22:00 hr. The black solid lines represent the location and orientation angle of the wind turbines that are yawed into the direction of the mean incoming velocity. Such contour plots are generated by accessing the database as illustrated in analysis code snippets in Appendix C.}
\label{fig:dirunal_inst_u}
\end{figure}

Next, snapshots of instantaneous velocity and potential temperature fields at two representative day and night times (15:00 hr and 22:00 hr) are analyzed in more detail. Recall that  turbines  directly face the incoming averaged flow, and the wakes respond  accordingly shifting in the direction of the incoming wind. As exhibited in Fig. \ref{fig:dirunal_inst_u}a, at 15:00 hr under strong unstable conditions, the wake flow behind the wind turbines exhibit strongly meandering patterns around the streamwise  ($x$)-direction. As seen in Fig. \ref{fig:diurnal_velocity_profiles}b the precursor-wind farm simulation at 15:00 hr has a mean wind direction purely in the $x$ direction at hub height (thus the imposed yaw angle of all wind turbines is zero).  Later at 18:00 hr, as was shown in  Figs.
\ref{fig:diurnal_layout}b and  \ref{fig:diurnal_velocity_profiles}b     the incoming flow has a negative angle relative to the $x$-direction and the turbines yaw accordingly 
causing the wake flow behind the wind turbines to tilt towards the negative y-direction. At night-time,  at 22:00 hr, Fig. \ref{fig:diurnal_velocity_profiles}b and \ref{fig:dirunal_inst_u}b show that the incoming flow makes a positive angle with the $x$ direction, thus the wakes tilt towards the positive y-direction. The incoming turbulence has changed significantly in that it is far more small-scale and unorganized at large scales due to the strong stable stratification and much thinner boundary layer.   

\begin{figure}[H]
\centering
\includegraphics[width=0.9\textwidth]{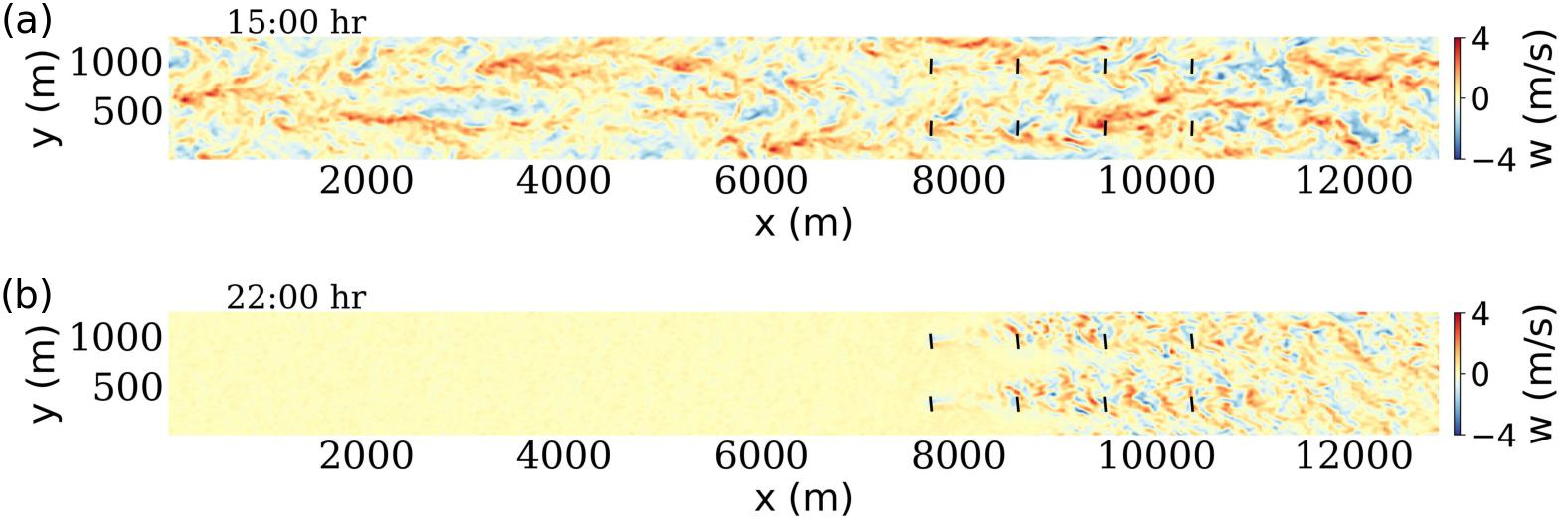}
\caption{Instantaneous vertical velocity at hub height  at (a) afternoon 15:00 hr, (b) evening 22:00 hr. The black solid lines represent the location and orientation angle of the wind turbines.}
\label{fig:dirunal_inst_w}
\end{figure}

Similar trends can be observed for the vertical velocity fluctuation contours, shown in  Fig. \ref{fig:dirunal_inst_w}. Under strong unstable conditions at \textcolor{black}{15:00 hr},  in Fig. Fig. \ref{fig:dirunal_inst_w}a strong vertical fluctuations are visible and they are enhanced in the wakes.   However, as the ground surface cools at night, the instantaneous vertical velocity fluctuations are greatly decreased in the incoming flow but becoming substantial only behind the wind turbines due to the increased TKE in the wakes.

\section{Analysis of thermal fields across evening transition}
\label{sec:resultstemp} 

Several observational studies \cite{zhou2012impacts,rajewski2013crop,smith2013situ} have also observed   night-time warming effects in the presence of wind farms. They report about $O(1\,K)$ rise at the ground surface, with variations depending on vegetation type and season. Prior simulation \cite{lu2011large} has also shown such an effect. Using the detailed dataset from our diurnal cycle simulation, we can provide additional confirmation and physical insights into the associated phenomena.

\begin{figure}[H]
\centering
\includegraphics[width=0.9\textwidth]{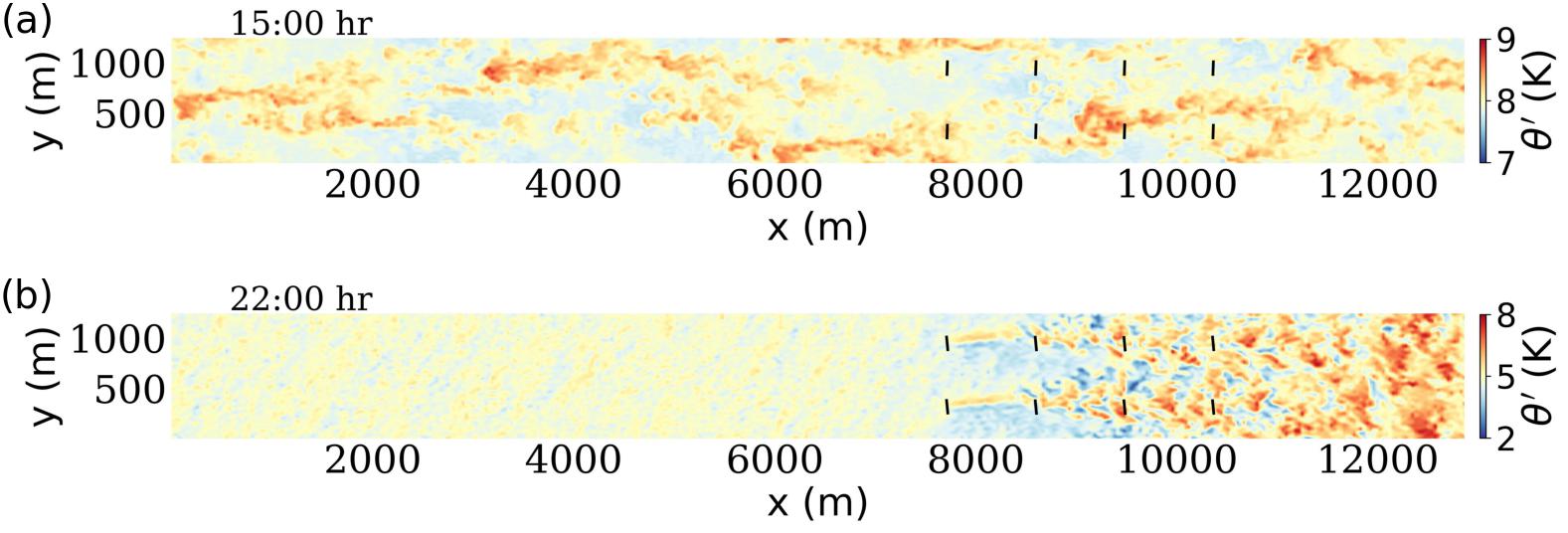}
\caption{Instantaneous temperature at hub height recorded in the (a) afternoon 15:00 hr, (b) night-time, 22:00 hr.}
\label{fig:dirunal_inst_T}
\end{figure}

\begin{figure}[H]
\centering
\includegraphics[width=0.9\textwidth]{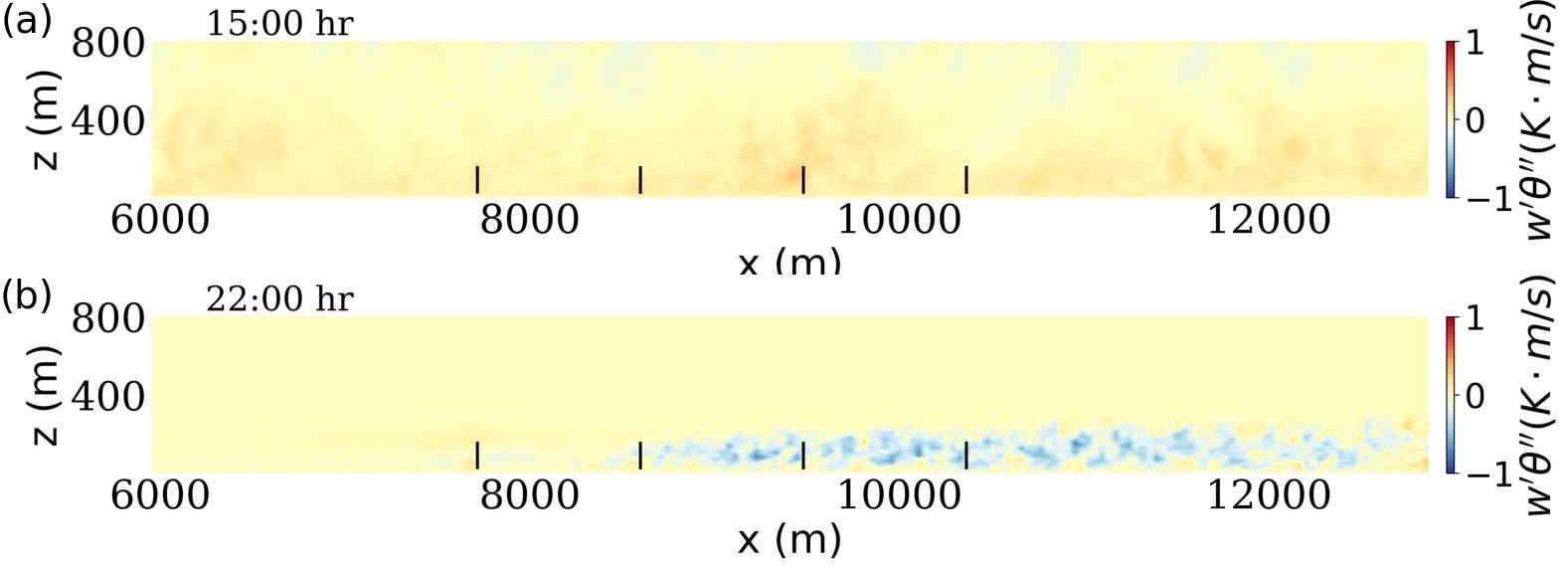}
\caption{Side view contours of spanwise averaged vertical turbulent heat flux within the wind farm domain in the (a) afternoon, 15:00 hr. (b) night-time 22:00 hr. The black solid lines represent the location of the wind turbines.}
\label{fig:dirunal_inst_heatflux}
\end{figure}

Fig. \ref{fig:dirunal_inst_T} shows   instantaneous potential temperature contours exhibiting significant spatial fluctuations during the strong convective conditions at 15:00 hr. For night-time conditions, e.g. at 22:00 hr, as the ground surface is cooling,  the potential temperature fluctuations at the inlet become smaller amplitude and occur at much smaller scales. A significant effect can be seen behind the first row of turbines in which there is temperature reduction at hub height while behind the last row there is warming on average. This warming effect is driven by the enhanced turbulent mixing caused by the wind farm, resulting in a substantial negative vertical turbulent heat flux. This flux brings warmer air from higher elevations down to cooler, lower elevations, as shown in Fig. \ref{fig:dirunal_inst_heatflux}b.  Conversely, during the day, radiative heating at the surface leads to positive vertical turbulent heat flux, as shown in Fig. \ref{fig:dirunal_inst_heatflux}a but no \textcolor{black}{discernible} change in average temperature in the wind farm wake. Fluxes shown in Fig. \ref{fig:dirunal_inst_heatflux} are computed based only on spanwise averaging and thus contain significant remaining fluctuations.

\begin{figure}[H]
\centering
\includegraphics[width=0.9\textwidth]{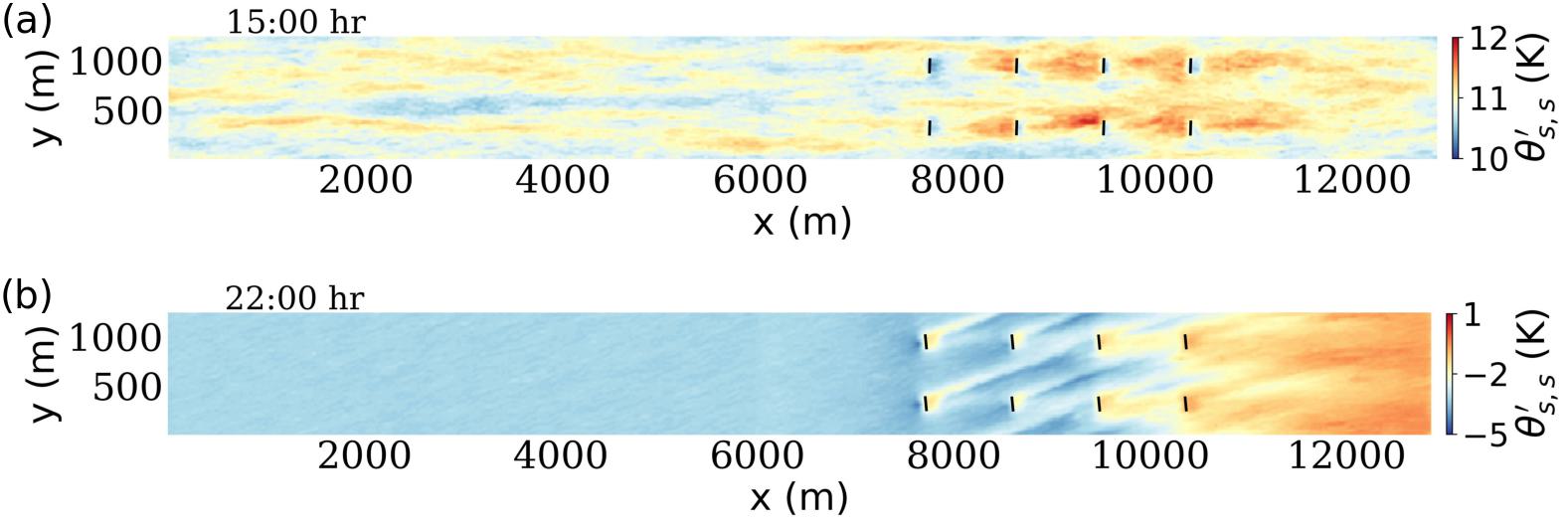}
\caption{Instantaneous surface soil temperature recorded at (a) afternoon 15:00 hr. (b) evening 22:00 hr.}
\label{fig:dirunal_inst_ST}
\end{figure}

Fig. \ref{fig:dirunal_inst_ST} shows the temperature at the ground surface. Clearly,  the wind farm wake flow has a significant footprint on the ground surface temperature. At 15:00 hr, the wake flow slows down behind the wind turbines, reducing the turbulent heat flux into the air due to a smaller friction velocity, as indicated by equation \ref{eq:air turbulent heat flux}. This results in a warming region directly behind the turbines due to increased heat flux partitioning from radiation into the ground, corresponding to the wake region in the air flow. At night-time at 22:00 hr, the aforementioned warming effect of the air flow also are observed to extend to the soil, increasing ground surface temperature behind the wind farm. Notably, the direction of the wake visible in the ground temperature distribution differs from that in the air at hub-height, as seen when comparing Fig. Fig. \ref{fig:dirunal_inst_T}b and Fig. \ref{fig:dirunal_inst_ST}b. This difference is caused by the substantial veer effect significantly changing the wind direction at different elevations during nighttime, This is also visible in Fig. \ref{fig:diurnal_velocity_profiles}b.

We conclude that the present simulations confirm prior observations of surface heating downstream of wind farms during night-time operations under strongly stable stratified conditions. The mechanism is relatively clear: the much thinner vertical extent of the boundary layer is very sensitive to additional mixing due to wind turbine wake turbulence, which brings hotter fluid closer to the surface thus raising the temperature in the lower parts of the boundary layer including the ground surface. We remark that if we had imposed a spatially uniform surface heat flux or imposed temperature at the ground, the strong spatial variability we observe in thermal fields with the physically more appropriate boundary condition would have been masked. 

\section{Analysis of power production across morning transition}
\label{sec:resultspower} 

 The power for each turbine at all times is available as part of the database and can be queried as illustrated in the code snippet presented in Appendix C. 
Fig. \ref{fig:dirunal_power}a shows the power of two individual turbines in the front (turbine 1) and back (turbine 7) rows, during the initial 100 seconds of the diurnal cycle, starting from 15:00 hr.  The signals show fluctuations occurring at time-scales of about O(15) seconds. With a mean velocity of O(10) m/s, this corresponds to turbulence length-scales of about 150 m, consistent with averaging them over the rotor disk of diameter O(120) m. Conversely, the signals also display periodic oscillations at a time-scale of 2 seconds, which is consistent with a blade passage frequency (3$\times$ rotor frequency). We have verified that the peaks in these signals correspond to two blades being in the upper parts of their cycle, exposed to faster winds. 

We next document the power production of the entire wind farm across the full diurnal cycle in Fig. \ref{fig:dirunal_power}b.
Significant variations in generated power can be observed during the 24 hour period, even though the imposed geostrophic wind velocity is kept constant at 15 m/s. Production rises during the day and reaches a peak late afternoon near \textcolor{black}{17:00 hr}, then quickly decreases to increase again during the first part of night-time. After midnight, a slow decrease is observed again. Also visible are larger power fluctuation levels during the day-time unstable atmospheric conditions. 

\begin{figure}[H]
\centering
\includegraphics[width=0.8\textwidth]{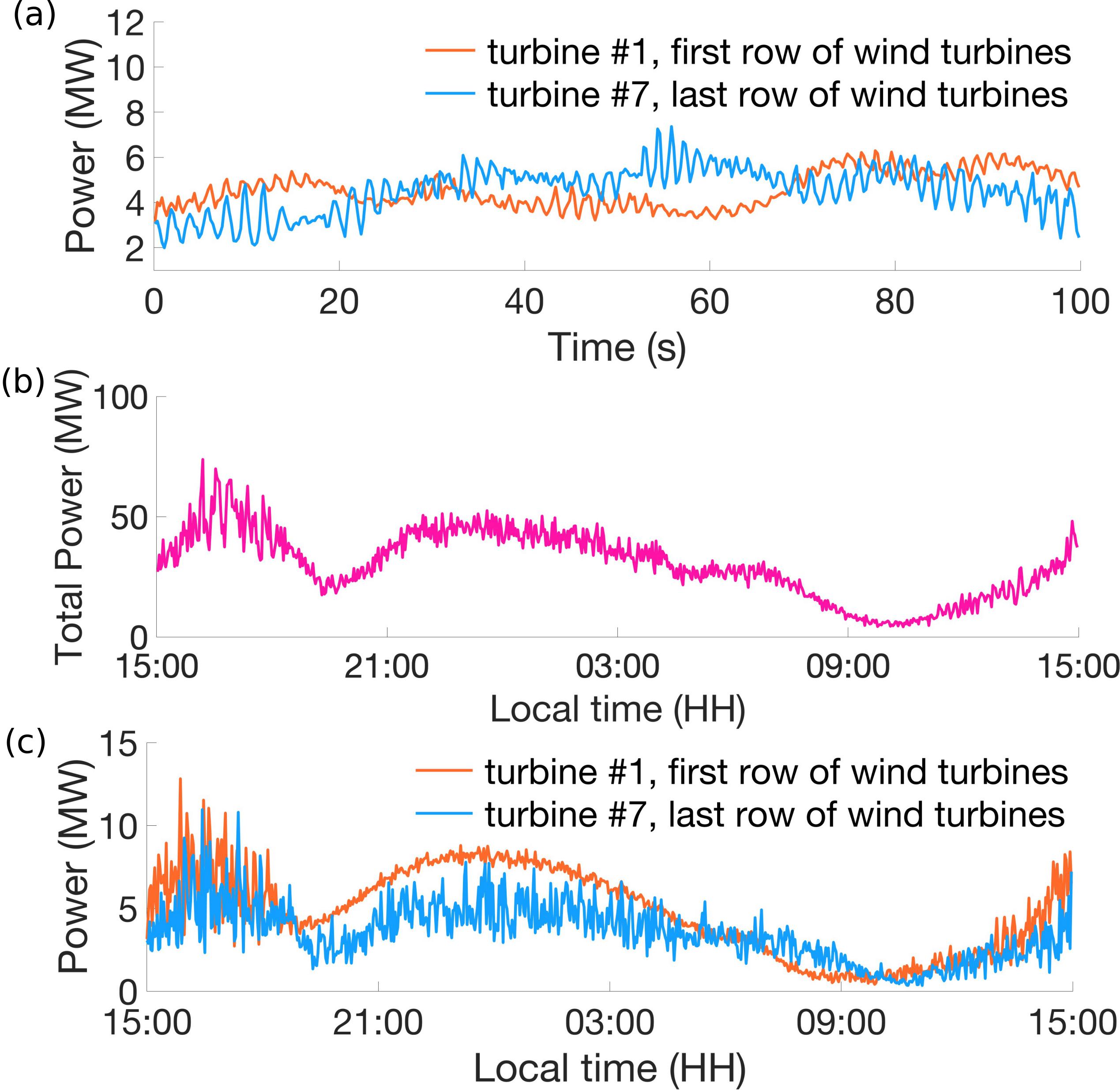}
\caption{(a) Power time series for turbine \#1 (front row) and turbine \#7 (back row) over the initial 100 seconds of the data, starting from 15:00 hr,  (b) Power time series of power generated by entire wind farm across the diurnal cycle (sum of each turbine's generated power), and (c) Power time series for turbine \#1 (front row) and turbine \#7 (back row) over the entire diurnal cycle. The generated power for each turbine at all times is available as part of the database (aerodynamic torque multiplied by the instantaneous rotor angular velocity), and can be queried as illustrated in the code snippet presented in Appendix C.}
\label{fig:dirunal_power}
\end{figure}

\textcolor{black}{Focusing} now on individual turbines in the front (turbine 1) and back (turbine 7) rows during the entire cycle, Fig. \ref{fig:dirunal_power}c displays their generated power.  Comparing the different behaviors of turbines in the front and back rows, we recall that turbines in the first row in a wind farm typically experience stronger wind velocities compared to those located downstream that are affected by wakes. As shown in Fig. \ref{fig:dirunal_power}c this pattern is indeed observed throughout most of the diurnal cycle as turbine 1 generates significantly more power than turbine 7. However, during the morning transition between \textcolor{black}{06:00 hr} and \textcolor{black}{10:00 hr}, an unusual phenomenon occurs, where the power generated by turbine 7 in the last row is actually higher than that of turbine 1 in the first row. The morning transition period is of great interest because it marks the shift from a stable to a convective boundary layer. During this time, the stability of the atmosphere rapidly changes, which impacts wind shear, turbulence, and overall wind flow. For wind energy, this transition is particularly important because the changing turbulence and wind speed profiles can affect turbine performance and energy generation. 

\begin{figure}[H]
\centering
\includegraphics[width=0.6\textwidth]{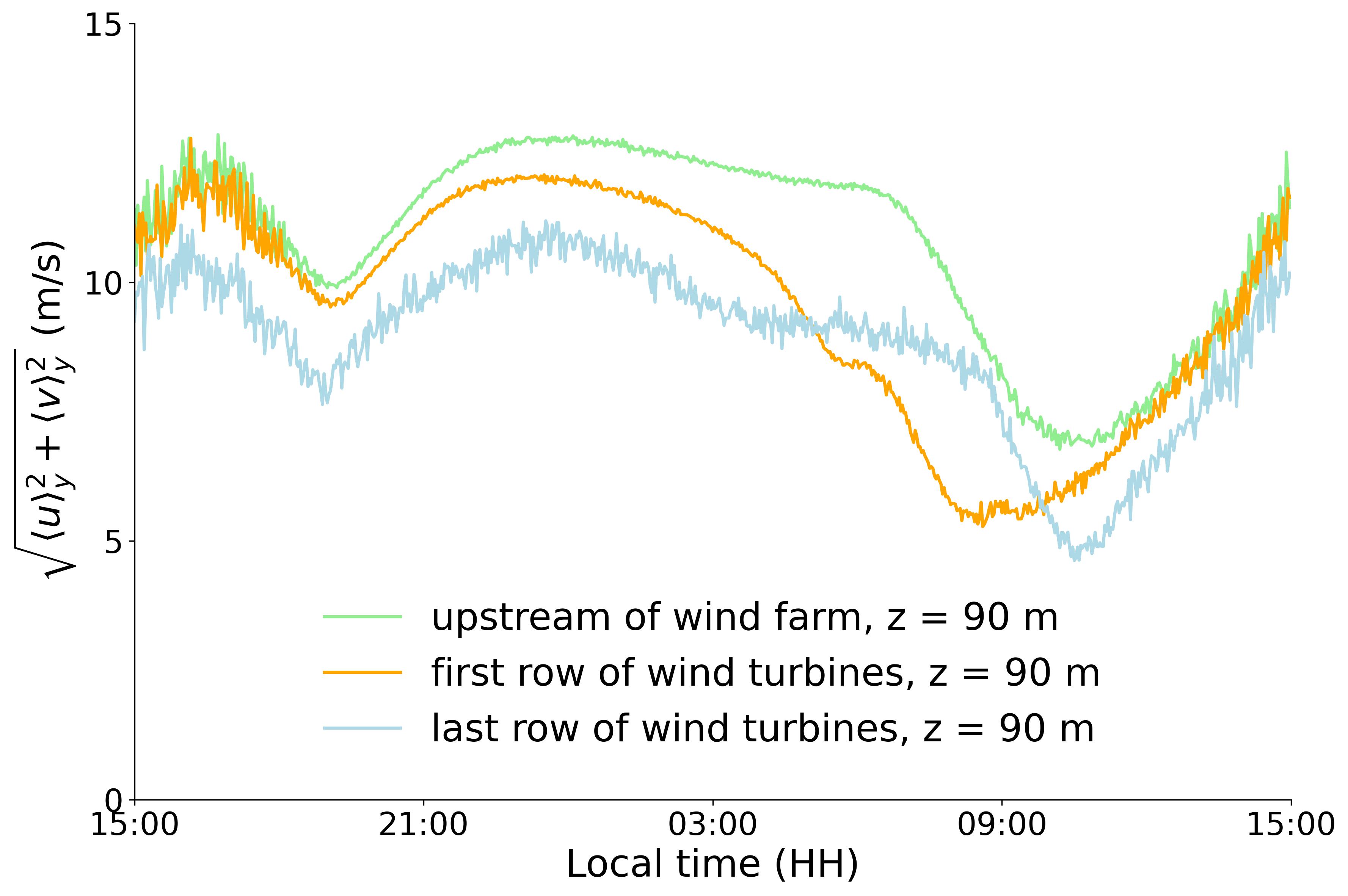}
\caption{Time evolution of the spanwise-averaged velocity during entire diurnal cycle, in the precursor domain (green line), directly in front of the first row of wind turbines (orange line), and directly in front of the last row of wind turbines over a diurnal cycle (blue line).}
\label{fig:diurnal_mean_vel_farm}
\end{figure}

We now present morning transition data for velocity, temperature and turbulent kinetic energy distributions to shed light onto the origin of the observation of larger power production of downstream turbines during parts of the cycle. Figure \ref{fig:diurnal_mean_vel_farm} shows the hub-height velocity as function of time during the diurnal cycle at three streamwise locations: in the precursor (upstream of the wind farm), at the location of the first row and at the last row. These results are consistent with the power signals in the sense that roughly between \textcolor{black}{06:00 hr} and \textcolor{black}{09:00 hr} the hub-height wind velocity is larger for the last row compared to the first row. We also observe that the difference between the upstream and first-row velocity significantly increases during the night time, especially after \textcolor{black}{03:00 hr}. This effect is caused by increased flow blockage and deflection of the low-level jet.

\begin{figure}[H]
\centering
\includegraphics[width=0.9\textwidth]{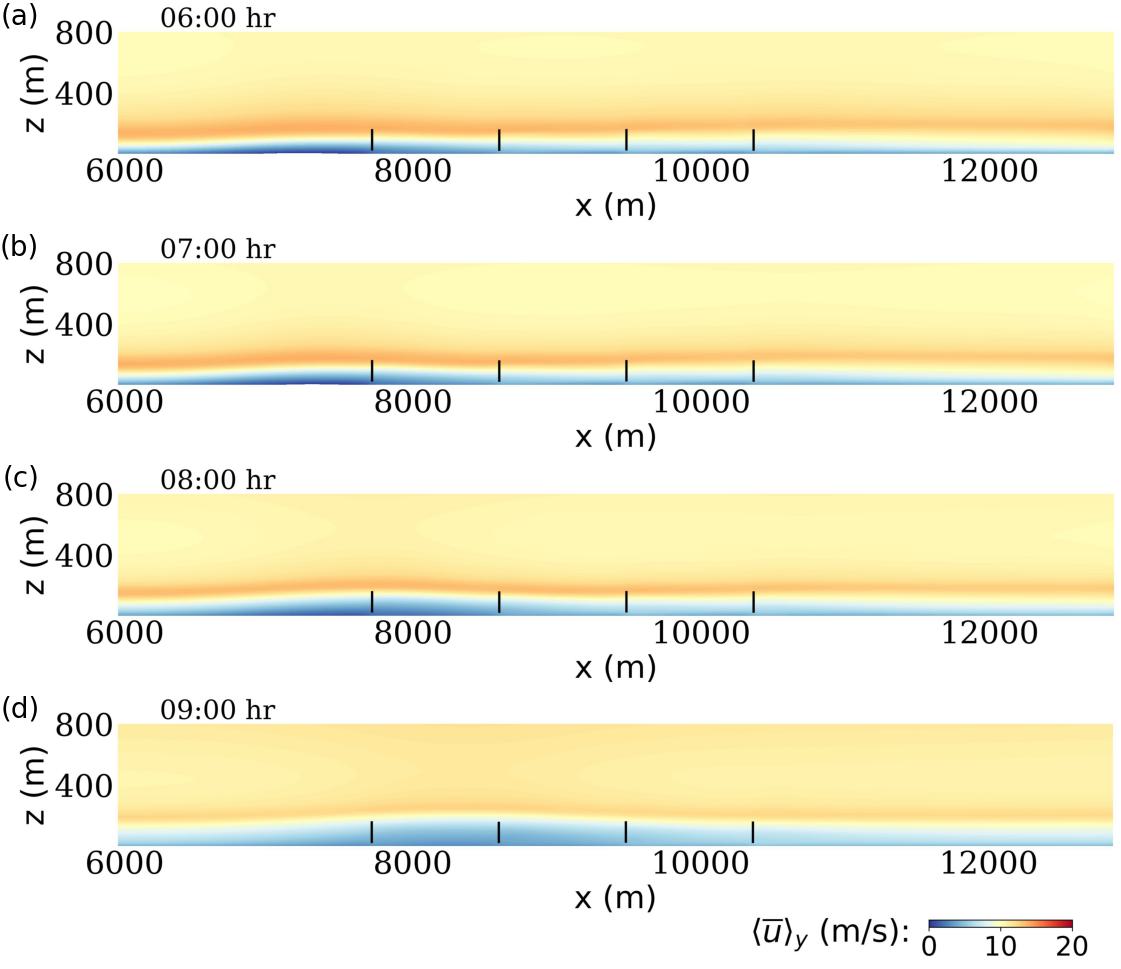}
\caption{Contours of the mean, 10-minute and spanwise-averaged streamwise velocity within the wind farm domain on $x-z$ planes, recorded in the (a) early morning 06:00 hr, (b) early morning 07:00 hr, (c)   morning 08:00 hr, and (d)  morning 09:00 hr. The black solid lines represent the locations of the wind turbines.}
\label{fig:dirunal_inst_u_side_06000900}
\end{figure}

The transition from stable to convective occurs gradually between approximately \textcolor{black}{06:00 hr} and 09:00 hr. During this period, a strong \textcolor{black}{nocturnal low-level jet is observed} above the ground, as shown in Fig. \ref{fig:diurnal_velocity_profiles}a and Fig. \ref{fig:dirunal_inst_u_side_06000900}, characterized by a narrow band of high velocity just above hub height. Initially, the incoming flow features a relatively broader band of low-level jets with less turbulence. As the flow approaches the wind farm, particularly within the first few rows of turbines, it rises to a higher elevation. 

\begin{figure}[H]
\centering
\includegraphics[width=0.9\textwidth]{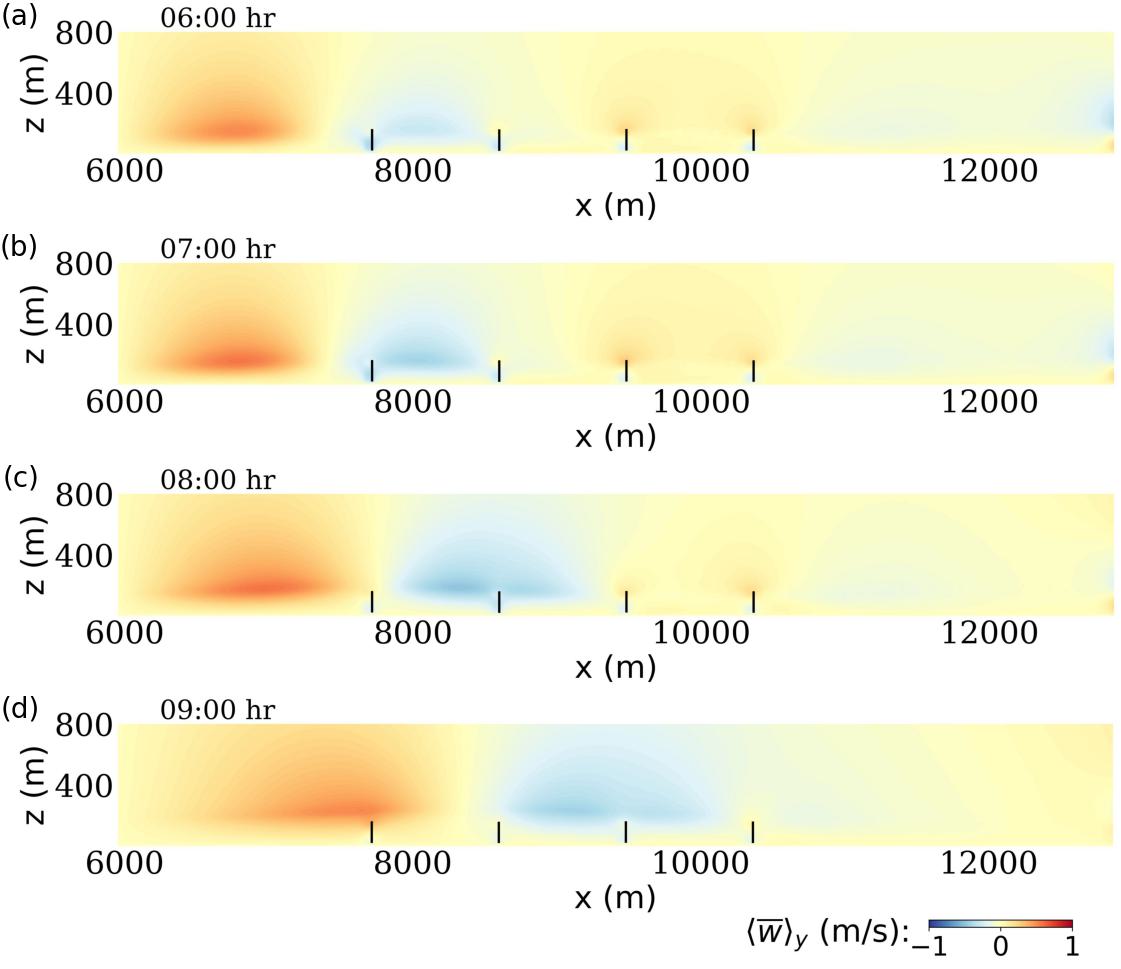}
\caption{Contours of the mean, 10-minute and spanwise-averaged vertical velocity within the wind farm domain on $x-z$ planes, recorded in the (a) early morning 06:00 hr, (b) early morning 07:00 hr, (c)   morning 08:00 hr, and (d)  morning 09:00 hr. The black solid lines represent the location of the wind turbines.}
\label{fig:dirunal_inst_w_side_06000900}
\end{figure}

\begin{figure}[H]
\centering
\includegraphics[width=0.9\textwidth]{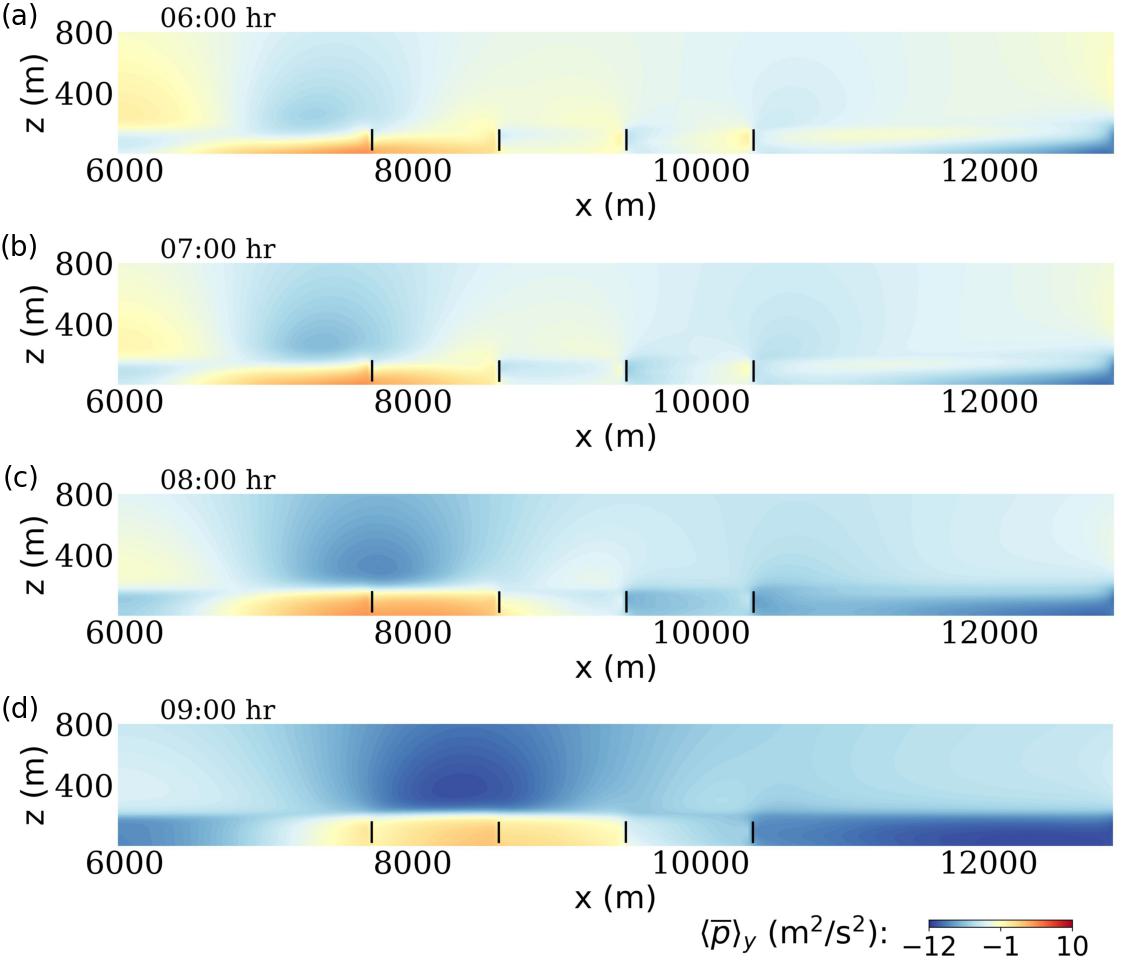}
\caption{Contours of the mean, 10-minute and spanwise-averaged pressure within the wind farm domain from side view recorded in the (a) early morning 06:00 hr, (b) early morning 07:00 hr, (c)   morning 08:00 hr, and (d)  morning 09:00 hr. The black solid lines represent the location of the wind turbines.}
\label{fig:dirunal_inst_p_side_06000900}
\end{figure}

The lifting of the \textcolor{black}{ nocturnal}  low-level jet  as it encounters the wind farm is caused by the blockage effect of the turbines, which becomes quite significant when the boundary layer height is comparable to the turbine height. This blockage effect is characterized by a strong induced positive vertical velocity just in front of the wind farm, along with a pronounced adverse pressure gradient and a negative vertical pressure gradient, as exhibited in Fig. \ref{fig:dirunal_inst_w_side_06000900} and Fig. \ref{fig:dirunal_inst_p_side_06000900}. 

The pressure distribution within the wind farm, shown in Fig. \ref{fig:dirunal_inst_p_side_06000900}, shows the expected high pressure region in front of the wind farm which acts to deflect the flow upwards, as seen in Fig. \ref{fig:dirunal_inst_w_side_06000900}. Interestingly, the effect propagates slightly downstream at \textcolor{black}{09:00 hr} consistent with the vertical velocity distribution in Fig. \ref{fig:dirunal_inst_w_side_06000900} that also shows the downward moving regions shifting downstream in the farm at later times.

\begin{figure}[H]
\centering
\includegraphics[width=0.6\textwidth]{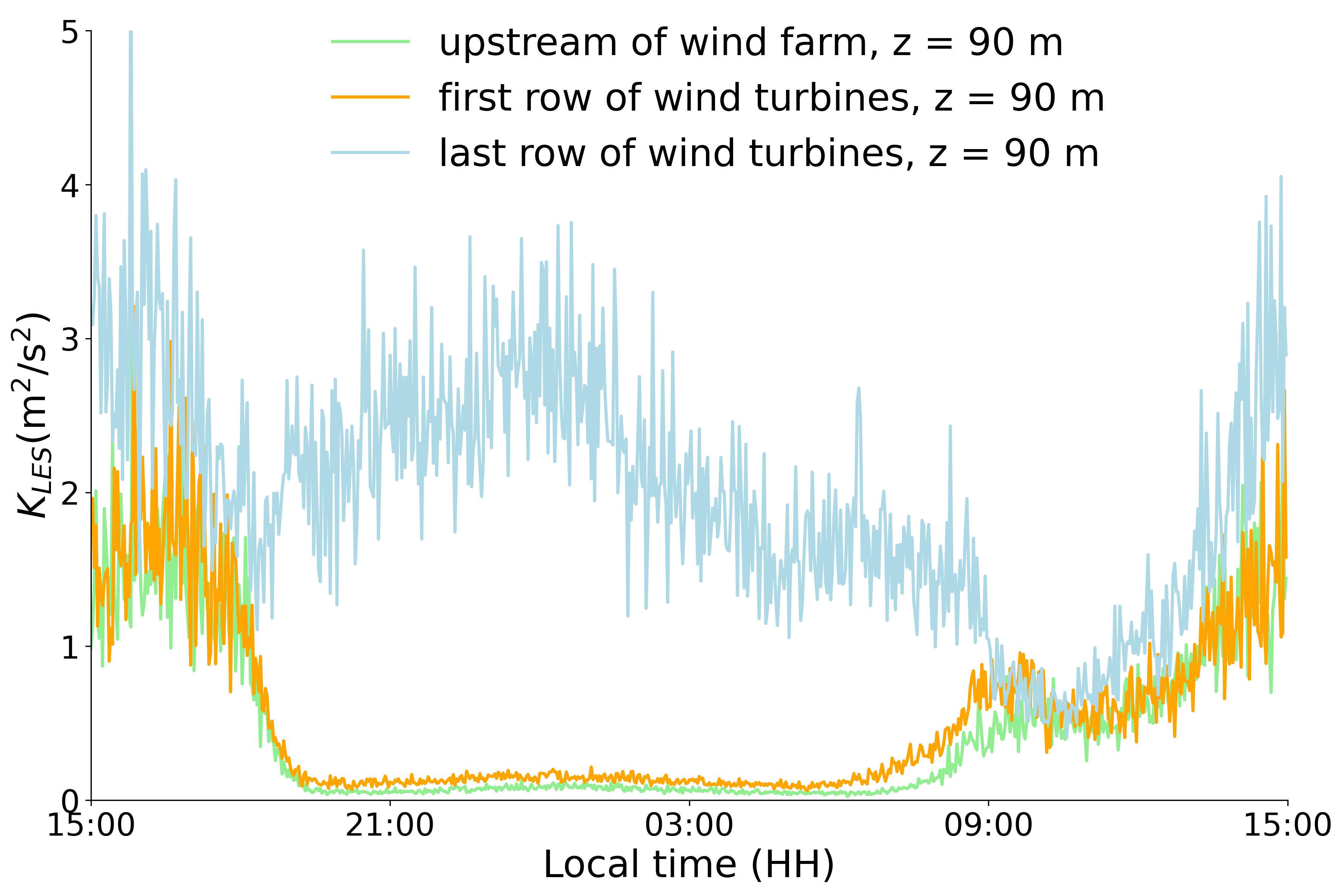}
\caption{Time evolution of the spanwise-averaged turbulent kinetic energy in the precursor domain (green line), directly in front of the first row of wind turbines (orange line), and directly in front of the last row of wind turbines (blue line) over a diurnal cycle.}
\label{fig:diurnal_mean_turbulence_farm}
\end{figure}

Thus, the rapid decrease in velocity seen by the first row of turbines is associated with flow blockage and low-level jet deflection. How about the last row's velocity that does not decrease accordingly? To answer this question we examine the turbulence levels, shown in Fig. \ref{fig:diurnal_mean_turbulence_farm} across the diurnal cycle. As is expected, the turbulence level at the last row of wind turbines is consistently stronger than at the first row,  \textcolor{black}{especially} during night-time stable conditions and the morning transition. Even during very stable conditions the turbine wakes maintain more elevated levels of turbulence as compared to the much decreased turbulence levels of the incoming stably stratified flow. 
 
\begin{figure}[H]
\centering
\includegraphics[width=0.8\textwidth]{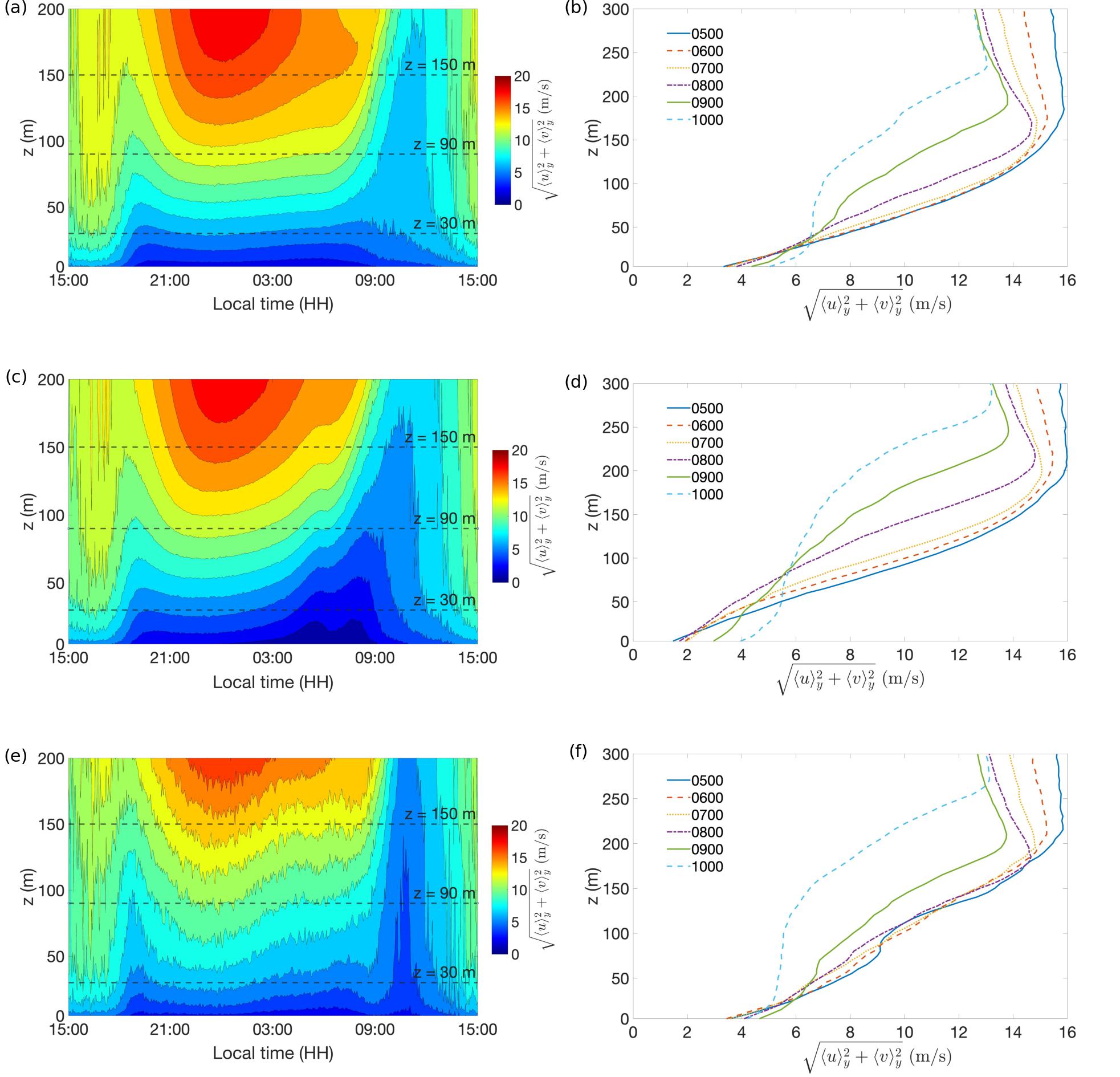}
\caption{Contour plots (left column) and vertical profiles  (right column) of mean velocity magnitude over a diurnal cycle. (a), (c), and (e) are for averages in the precursor domain, directly in front of the first row of wind turbines, and directly in front of the last row of wind turbines, respectively. (b), (d), and (f) are  profile plots of spanwise-averaged velocity magnitude from 05:00 hr to 10:00 hr in the precursor domain, directly in front of the first row of wind turbines, and directly in front of the last row of wind turbines, respectively.}
\label{fig:diurnal_mean_velocity_location}
\end{figure}

In order to identify the effects of the increased turbulence levels on the mean velocity distributions arriving at the turbine hub-height, it is instructive to analyze the diurnal cycle time evolution of mean velocity across heights. This is shown in  Fig. \ref{fig:diurnal_mean_velocity_location}, both as contour plots in the lower 200 m across the full diurnal cycle  and vertical profiles at select times encompassing the morning transition. The blockage effect is visible in comparing figures (a,b) to (c,d), the latter appearing shifted upwards. 
The effect of increased mixing during the \textcolor{black}{06:00 hr - 09:00 hr} time period of interest can be seen as the ``flattening out'' of the velocity iso-contours during that time-frame when comparing (c,d) with panels (e,g). In particular iso-contours are more separated, hence the flow is better mixed in panels (e,f) between times \textcolor{black}{06:00 hr - 09:00 hr}  at elevations between $z = 30\,\text{m}$ and $z = 150\,\text{m}$, which corresponds to the range of heights the turbines can reach.   During the morning transition, the boundary layer height is at its lowest, as shown in Fig. \ref{fig:diurnal_stress_ABL}. As a consequence, the low-level jets exhibit the largest vertical gradient at this time compared to other periods. Moreover, the low-level jet of the incoming flow is positioned just above the wind turbines. In the presence of the turbines, particularly at the last row, strong turbulence mixing is triggered, bringing high-velocity momentum from the higher elevations within the narrow band of the low-level jet downward into the region swept by the turbines, resulting in higher wind speeds experienced by the turbines. 

We conclude that the unexpected behavior of larger power in the last row of turbines compared to the first row can be ascribed to a combination of two effects: reduced power in the front row due to increased blockage during night-time and increased mixing of the deflected low-level jet into the wind farm due to wind farm turbulence which remains relatively strong even at \textcolor{black}{night-time}.  

\section{Conclusions}
\label{sec:conclusions}

Data generated from LES of a wind farm placed in a diurnal cycle of atmospheric boundary layer evolution were analyzed. Placement of the simulation data into a web-services accessible database system facilitated detailed analysis to address specific research question. Details about the JHTDB-wind database system and its access methods are provided in the appendices. Data analysis focused on two features of the complex dynamics of atmosphere-wind farm interactions: The thermal fields during the evening transition and a power production anomaly observed during the morning transition. Results highlighted the importance of using a soil conduction model  to avoid having to specify either prescribed surface temperature or heat flux which would have eliminated the strong spatial variability ultimately observed in these quantities inside and downstream of the wind farm.  

Results confirmed that during night-time when the ABL is strongly stable and the boundary layer is very shallow (of similar height to the wind turbines), increases in surface temperature of several degrees are observed. Analysis of the velocity and potential temperature fields during the evening transition show that increased turbulence levels in the wind farm wake cause warmer air from above (the residual layer) to be entrained downward causing local increases in temperature. 

Another unusual  behavior observed in the simulation was a higher power production for downstream turbines as compared to the front turbines during the morning transition. This phenomenon too was traced to increased turbulence levels in the wind farm, which had a disproportionate effect during time periods with shallow incoming boundary layer and strong wind farm blockage effects, i.e. under strongly stable conditions. 

Many additional research questions can be posed based on the LES of a wind farm during a full 24 hour period. The availability of publicly accessible data consisting of all the relevant fields in 4 (space and time) dimensions as well a along all of the turbine blades that can be easily queried as part of JHTDB-wind should help facilitate such research and provide added value to high-fidelity simulations of representative wind farms. 

\vspace{1cm}

\noindent {\bf Data Availability:}

 The entirety of the wind farm data is available for open access at the JHTDB-wind website (\url{https://turbulence.idies.jhu.edu/datasets/windfarms}), and demonstration data access (DEMO) codes (\url{https://turbulence.idies.jhu.edu/database/wind}) are available on this site as well, as described in this paper (Appendix C).

\section*{Acknowledgments}
The authors are grateful to Dr. Tony Martinez-Tossas for his help with the actuator line modeling and to Michael Schnaubelt  and other IDIES staff for support in creation of the JHTDB-wind database.  We thank Dr. Ned Patton for insightful comments regarding the simulations and data, and to Prof. Ben Schafer for his steadfast support and encouragement of JHTDB-wind. The project was made possible by a seed grant from the Ralph O'Connor Sustainable Energy Institute Research Initiative (ROSEI) at JHU, and by NSF grant \# 2034111 as well a joint NSF-DOE grant \# 2401013. The JHTDB project is supported by NSF (CSSI-2103874) and the Institute for Data Intensive Engineering and Science (IDIES) and its staff.   We are grateful for the high-performance computing (HPC) resources and assistance received from both Cheyenne (doi:10.5065/D6RX99HX), made available by NCAR’s CISL and sponsored by the NSF, and the Advanced Research Computing at Hopkins (ARCH) core facility (rockfish.jhu.edu), supported by the NSF under grant OAC1920103.

 \section*{Appendix A: Stability correction functions} \label{sec:stabfunctions}
 
In this Appendix we list the  empirical stability correction functions implemented in LESGO, based on Refs. \cite{Chenge_Brutsaert_2005,Brutsaert_2005}.  For the momentum boundary conditions, $\phi_m(\zeta)$ under unstable ($\zeta<0$) and stable ($\zeta>0$) atmospheric conditions are given by:
 
\begin{align}
    \phi_m(\zeta) = 
\begin{cases}
\dfrac{a_u + b_u (-\zeta)^{4/3}}{a_u -\zeta} \quad & ~~  {\rm for}~~ -\zeta \leq b_u^{-3} \ \text{(unstable)}, \\
1.0 \quad & ~~  {\rm for}~~ -\zeta > b_u^{-3} \ \text{(unstable)}, \\
1 + a_s \dfrac{\zeta + \zeta^{b_s} (1 + \zeta^{b_s})^{-1 + \frac{1}{b_s}}}{\zeta + (1 + \zeta^{b_s})^{\frac{1}{b_s}}} & ~~ {\rm for}~~ \zeta  > 0 \mspace{22mu} \text{(stable)},
\end{cases}  \label{phim_func}
\end{align}
where the constants for unstable atmospheric conditions are $a_u=0.33$, $b_u=0.41$, and for stable conditions are $a_s=6.1$, $b_s=2.5$. The $\Psi_m(\zeta)$ stability correction functions resulting from evaluating Eq. \eqref{Psi_m_func} by using  Eq. \eqref{phim_func} are,
\begin{align}
\Psi_m(\zeta) = 
\begin{cases}
\ln(a_u -\zeta) - 3 b_u (-\zeta)^{1/3} \\ + \dfrac{b_u a_u^{1/3}}{2} \ln \left[ \dfrac{(1 + (-\zeta/a_u)^{1/3})^2}{(1 - (-\zeta/a_u)^{1/3} + (-\zeta/a_u)^{2/3})} \right] \\
\quad + 3^{1/2} b_u a_u^{1/3} \tan^{-1} \left[ (2(-\zeta/a_u)^{1/3} - 1) / 3^{1/2} \right] + \Psi_0 \quad & ~~  {\rm for}~~  -\zeta \leq b_u^{-3} \ \text{(unstable)}, \\
\Psi_m(b_u^{-3}) \quad & ~~  {\rm for}~~  -\zeta > b_u^{-3} \ \text{(unstable)},\\ 
-a_s \, \ln \left( \zeta + (1 + \zeta^{b_s})^{\frac{1}{b_s}} \right) &  ~~  {\rm for}~~ \zeta  > 0 \mspace{22mu} \text{(stable)},
\end{cases}
\end{align}
where, the integration constant $\Psi_0 = -\ln a_u +3^{1/2}b_u a_u^{1/3}\pi/6$.

For the scalar potential temperature field  \cite{Chenge_Brutsaert_2005, Brutsaert_2005}, $\phi_h(\zeta)$ for unstable and stable atmospheric conditions are given by,
\begin{align}
    \phi_h(\zeta) = 
\begin{cases} 
\dfrac{c + d (-\zeta)^n}{c + (-\zeta)^n} & ~~~~~~~  {\rm for}~~~ \zeta  < 0 \ \text{(unstable)}, \\
&\\
1 + a_h \left[\dfrac{\zeta + \zeta^{b_h} (1 + \zeta^{b_h})^{-1 + \frac{1}{b_h}}}{\zeta + (1 + \zeta^{b_h})^{\frac{1}{b_h}}}\right] & ~~~~~~~  {\rm for}~~~ \zeta  > 0 \ \text{(stable)},
\end{cases}  \label{phi_h}
\end{align}
where the constants are $a_h = 5.3$, $b_h = 1.1$, $c = 0.33$, $d = 0.057$, and $n = 0.78$.
Substituting Eq. \eqref{phi_h} in Eq. \eqref{psi_h_func} gives,
\begin{align}
    \Psi_h(\zeta) = 
\begin{cases} 
\dfrac{(1 - d)}{n} \ln \left[ \dfrac{(c + (-\zeta)^n)}{c} \right] & ~~~~~~~  {\rm for}~~~ \zeta  < 0 \ \text{(unstable)}, \\
-a_h \, \ln \left( \zeta + (1 + \zeta^{b_h})^{\frac{1}{b_h}} \right) & ~~~~~~~  {\rm for}~~~ \zeta  > 0 \ \text{(stable)}. 
\end{cases}  \label{psi_h_func}
\end{align}

\section*{Appendix B: JHTDB-wind database construction} \label{sec:appstorage}
The LES data from the diurnal cycle simulation ingested into the database consist of three types:  (i) 3D field data across the simulation domains (precursor and wind farm domains), (ii) wind turbine data for each wind turbine from the integration of quantities along each of the blades for each turbine from the actuator line model, and (iii) detailed rotor blade data from the actuator line model along each of the rotor blades in the wind farm.

\subsection*{B1: 3D field data}
As shown in Table \ref{tab:tabledomain}, the fine resolution LES of the two combined domains involved  $N_x \times N_y \times N_z = (432+432) \times 144 \times 480$ grid points, with $\Delta x = 18.375 \ \text{m}, \Delta y = 8.75 \ \text{m}, \Delta z = 5 \ \text{m}$.   In order to reduce size of stored data,  the output has been spectrally filtered  in horizontal planes and sampled every other point horizontally. Also, fringe regions have been excluded prior to merging the data in the precursor and wind farm domains. The fringe regions excluded in the precursor domain amount to 2/8 of the domain in the $x$ direction and to 1/8 in the wind farm domain. Therefore,  the  size of the finally stored output files at each time step is $N^s_x \times N^s_y \times N^s_z = 351 \times 72 \times 480$, since $N^s_y=N_y/2$ and $N^s_x = (N_x/2)\times 6/8 + (N_x/2)\times 7/8$. 

 Field data are stored at every 10 time steps of the LES. With the conservatively small time step used in the LES ($\Delta t_{\rm les} = 0.05$s), the CFL condition is easily met. The stored time-step ($\Delta t_{\rm db} = 0.5$s) also means that a fluid particle will move at most 7m in the streamwise direction if moving at the largest fluid velocity (geostrophic speed), i.e. less than the horizontal spacing of the stored data. Rotor speeds are higher and will move across several vertical grid-spacings. However, the corresponding rotor force field is very smooth (Gaussian filtered at scale $\epsilon=19\,m\approx 2(\Delta x\Delta y\Delta z)^{1/3}$) and hence the storage frequency of $\Delta t_{\rm db} = 0.5$s is still appropriate. At every time, there are six spatial fields stored: the three velocity components  $u(x,y,z,t)$, $v(x,y,z,t)$, and $w(x,y,z,t)$, the potential temperature field in the air $\theta(x,y,z,t)$, 
the (kinematic) pressure field $p(x,y,z,t)/\rho = p^*-u_ku_k/2$ (the SGS stress trace is not available and anyhow negligible), 
and the subgrid-scale eddy viscosity $\nu_{\rm SGS}(x,y,z,t)$ (computed dynamically using the Lagrangian scale dependent dynamic Smagorinksy model). In addition, there is the soil temperature field $\theta_s(x,y,z,t)$  stored on 31 vertical points, i.e. on $351 \times 72 \times 31$ grid points at every time. And, the wind turbine force field $f_x(x,y,z,t)$, $f_y(x,y,z,t)$, and $f_z(x,y,z,t)$ is also stored. The turbine force data is mainly zero in most of the domain. Hence, stored data includes data only only up to 200 meters since at higher positions the turbine forces are zero (they are also zero between turbines but for storage and coding simplicity the data there are still stored as zeroes).  The force field is not spatially filtered and subsampled, and is instead stored at the original LES resolution. As function of time, over the 24 hours (86{,}400 seconds) there are $172{,}800$ fields stored. 
 
Table \ref{tab:table3dfields} indicates the stored field variables, their units,  the corresponding spatial domain sizes, and the number of time-steps stored. 
 
\begin{table}[H]
\centering
\caption{Spatial data in JHTDB-wind, stored every 0.5 seconds}
\begin{tabular}{|c|c|c|c|c|}
\hline
Variable & name & units & \# grid-points stored & \# of time-steps stored \\ \hline
\hline
x-velocity $u$ & velocity & m/s & $351 \times 72 \times 480$ & $172{,}800$ \\ \hline 
y-velocity $v$ & velocity & m/s & $351 \times 72 \times 480$ & $172{,}800$ \\ \hline 
z-velocity $w$ & velocity & m/s & $351 \times 72 \times 480$ & $172{,}800$ \\ \hline 
potential temperature deviation $\theta^\prime$ & temperature &  K & $351 \times 72 \times 480$ & $172{,}800$ \\ \hline 
pressure (kinematic) $p$ & pressure&   $m^2/s^2$ & $351 \times 72 \times 480$ & $172{,}800$ \\ \hline 
SGS eddy viscosity $\nu_{\rm SGS}$ & eddyviscosity & $m^2/s$  & $351 \times 72 \times 480$ & $172{,}800$ \\ \hline 
soil temperature deviation $\theta^\prime_s$ & soiltemperature & K & $351 \times 72 \times 31$ & $172{,}800$ \\ \hline 
turbine force (kinematic) $f_x$ & force &  $m/s^2$ & $151 \times 144 \times 40 $ & $172{,}800$ \\ \hline
turbine force (kinematic) $f_y$ & force & $m/s^2$ & $151 \times 144 \times 40 $ & $172{,}800$ \\ \hline
turbine force (kinematic) $f_z$ &  force &$m/s^2$ & $151 \times 144 \times 40 $ & $172{,}800$ \\ \hline
\end{tabular}
\label{tab:table3dfields}
\end{table}
 
The fields are stored using the ZARR format \cite{Miles2023} with chunk sizes of $39 \times 72 \times 80$ for most quantities. Prior experience and tests with other JHTDB datasets have shown that ZARR chunk sizes of  $64^3$ provide optimal retrieval speeds and performance for typical data access modalities. Chunks should be large enough to accommodate  operations (interpolations, differentiation) that require data in a 3D neighborhood of a queried datapoint, while avoiding reading too large data segments into memory. The total amount of data stored for the diurnal cycle is about 40 Terabytes. 

\subsection*{B2: Data for each wind turbine}
For each of the 8 turbines, variables deduced from the actuator line modeling that represent turbine operation are recorded as function of time. \textcolor{black}{Unlike field data that are stored in kinematic units,   air density must be specified and   used to compute force and power data. A value of $\rho_{\rm air}=1.23$ kg/m$^3$ is used.} Wind turbine data are stored at the original simulation time-step, $\Delta t_{\rm les}=0.05$s. 

Unlike the 3D field data, these files are much smaller and require far less memory. Hence, they are stored as simple files using the Parquet format that is easy to query from various languages.  See Appendix B for examples. The turbine data stored consist of the   variables listed in table \ref{tab:tableturbinedata}: 

\begin{table}[H]
\centering
\caption{Turbine level variables, stored at the LES time resolution every 0.05 seconds}
\begin{tabular}{|c|c|c|c|c|c|c|}
\hline
No. & \makecell{Name of \\ variable} & \makecell{Name in \\ dataset} & Symbol & Unit & \makecell{Data size \\ $nt \times 2$} &\makecell{Data resolution \\ $\Delta t$ (s)}\\ \hline
1& Power & power &$P$ & W & \multirow{4}{*}{$172{,}800 \times 2$} & \multirow{4}{*}{0.05}\\ \hhline{|-----|~|}
2& Thrust force & thrust &$f_t$ & N &  & \\ \hhline{|-----|~|}
3& Rotor angular velocity & RotSpeed &$\omega$ & rad/s &  &  \\ \hhline{|-----|~|}
4& Turbine yaw angle & Yaw &$\gamma$ & rad &  &  \\ \hline
\end{tabular}
\label{tab:tableturbinedata}
\end{table}
 
 Similarly to getData functions to query 3D field data such as fluid velocity and pressure, the Turbine data can be queried using the {\it getTurbineData} calls from analysis programs (python, Matlab). 

\subsection*{B3: Wind turbine blade data along actuator line points}
For each of the blades ($8 \times 3 = 24$ blades), 
time histories of 19 different quantities (listed in Table \ref{tab:tableturbinebladedata}) are stored 
at 100 ALM collocation points. Similarly to turbine 
data, the blade data are stored as simple files using 
the Parquet format that is easy to query from various 
languages. The first column represents time in 
seconds, starting from $0\,s$ (corresponding to 15:00 hr) and going up to $86{,}400$s corresponding to \textcolor{black}{15:00 hr} of the next day), the second column is the 
turbine number, the third column is the blade number, and the remaining 100 columns are the values at each of the 100 actuator points from root to tip. Different quantities are 
stored in different files. 
They can be selected as 
{\it variable} specified in 
the
{getBladeData} calls, and are listed in table \ref{tab:tableturbinebladedata}.

\begin{table}[H]
\centering
\scriptsize
\caption{Local, blade‐level variables, stored at the LES time resolution every 0.05 s at each of the 100 blade ALM points (1–100)}
\begin{tabular}{|c|c|c|c|c|c|c|c|}
\hline

No. & \makecell{Name of \\ variable} & \makecell{Name in \\ dataset} & Symbol & Unit & \makecell{Data size \\$(nt \times 3) \times (n \ell +3) $ } &\makecell{Data resolution \\ $\Delta t \times \Delta \ell  $\\ (s $\times$ m)}   \\ \hline
1&ALM point x-position & xPos & $P_x$ &  \multirow{3}{*}{m} & \multirow{27}{*}{$(172{,}800\times 3) \times (100+3)$} & \multirow{27}{*}{$ 0.05 \times 0.615$}  \\ \hhline{|----|~|}
2&ALM point y-position & yPos & $P_y$ &  & &  \\ \hhline{|----|~|}
3&ALM point z-position & zPos & $P_z$ &  & &  \\ \hhline{|-----|~|}
4&\makecell{Perturbation velocity at\\ LES resolution, component 1 }& uy\_LES1 & $u'_{y,{\rm LES1}}$& \multirow{12}{*}{m/s} & &  \\ \hhline{|----|~|}
5&\makecell{Perturbation velocity at \\LES resolution, component 2} & uy\_LES2 & $u'_{y,{\rm LES2}}$&  & &  \\ \hhline{|----|~|}
6&\makecell{Perturbation velocity at \\optimal
resolution (0.25$c$), component 1} & uy\_opt1 & $u'_{y,{\rm opt}}$&   & &  \\ \hhline{|----|~|}
7&\makecell{Perturbation velocity at \\optimal
resolution (0.25$c$), component 2} & uy\_opt2 &$u'_{y,{\rm opt}}$& & &  \\ \hhline{|----|~|}
8&\makecell{Perturbation velocity correction \\ $u'_{y,{\rm opt}}-u'_{y,{\rm LES}}$, component 1 }& du1 & $\Delta u'_{y,{1}}$ &   & &  \\ \hhline{|----|~|}
9&\makecell{Perturbation velocity correction \\$u'_{y,{\rm opt}}-u'_{y,{\rm LES}}$, component 2 }& du2 & $\Delta u'_{y,{2}}$ &  & &  \\ \hhline{|-----|~|}
10&Angle of attack & alpha &$\alpha$& rad  & &  \\ \hhline{|-----|~|}
11&Lift coefficient & Cl &$C_L$& \multirow{2}{*}{-} & &  \\ \hhline{|----|~|}
12&Drag coefficient & Cd &$C_D$& & &  \\ \hhline{|-----|~|}
13&\makecell{Lift force per unit length}  & lift  &$F_L/\ell$ & \multirow{2}{*}{N/m} & &  \\ \hhline{|----|~|}
14&\makecell{Drag force per unit length}  & drag & $F_D/\ell$ &  & &  \\ \hhline{|-----|~|}
15&\makecell{Local relative velocity magnitude} & Vmag& $V_{mag}$ & \multirow{5}{*}{m/s} & &  \\ \hhline{|----|~|}
16&\makecell{Axial component of the local relative \\ velocity in blade-oriented coordinates}
& Vaxial & $V_{axi}$ &  & &  \\ \hhline{|----|~|}
17&\makecell{Tangential component of the local relative \\ velocity in blade-oriented coordinates} & Vtangential & $V_{tan}$&  & &  \\ \hhline{|-----|~|}
18&\makecell{Axial component of \\the local force (over segment $\Delta \ell$)} & axialForce & $F_{axi}$ & \multirow{4}{*}{N} & &  \\ \hhline{|----|~|}
19&\makecell{Tangential component of \\ the local force (over segment $\Delta \ell$)} & tangentialForce & $F_{tan}$ & & &  \\ \hline
\end{tabular}
\label{tab:tableturbinebladedata}
\end{table}

\vfill
\newpage

\section*{Appendix C: Sample JHTDB-wind access methods}
\label{sec:accessexamples}

\begin{figure}[H]
\centering
\includegraphics[width=\textwidth]{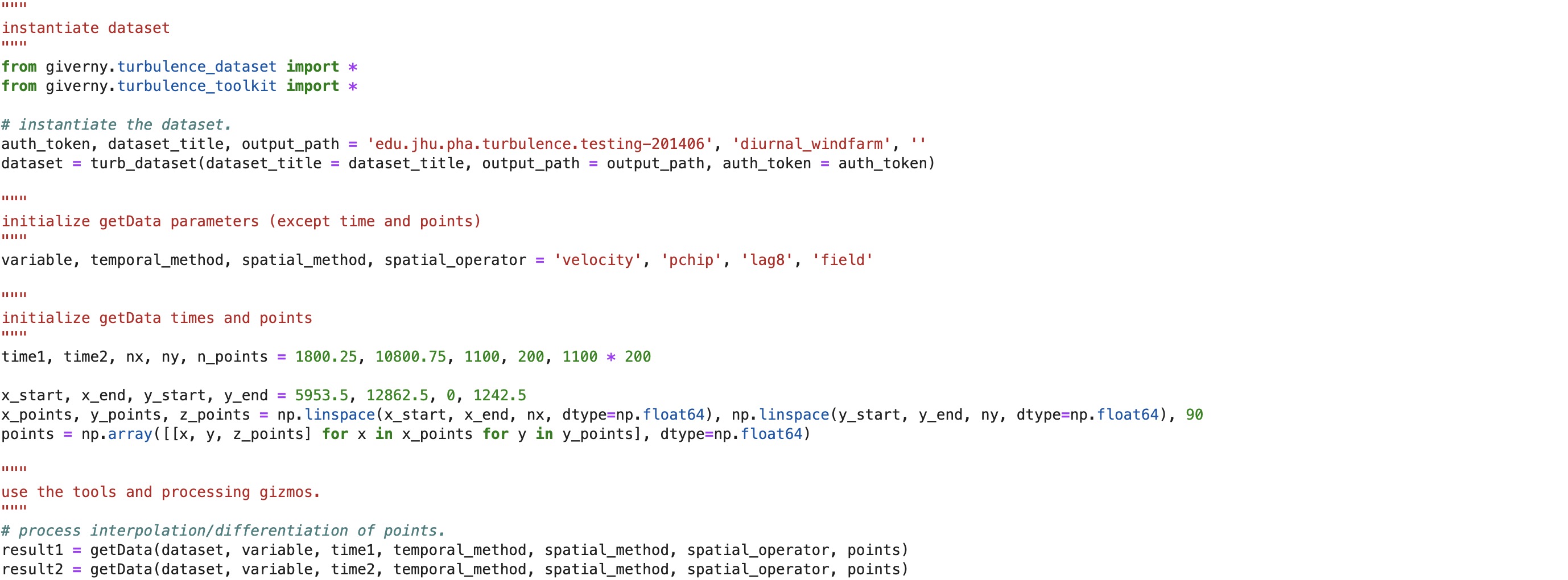}
\caption{Snippets of Python code that queries JHTDB-wind to obtain x-direction velocity snapshots on a horizontal plane at hub-height at two times $t=1{,}800.25$ s (about 15:30 hr) and $t=10{,}800.75$ s (about 18:00 hr).}
\label{fig:u_diurnal_demo_snippets}
\end{figure}

Similarly to JHTDB data, JHTDB-wind fields can be accessed via Web Services (\url{https://turbulence.idies.jhu.edu/database/wind}). Users can use familiar languages such as Python, Matlab, C or Fortran to execute analysis codes either remotely on their own computers or on a cloud service (SciServer) close to the data. Virtual sensors are placed in the flow field by means of the {\it getData(...)} function which requests field values at specified sets of points, time, etc.  Below (in Fig. \ref{fig:u_diurnal_demo_snippets}) we provide snippets of Python code that queries JHTDB-wind to obtain x-direction velocity snapshots on a horizontal plane at hub-height at two times, one at $t=1{,}800.25$ s (about 15:30 hr) and another at $t=10{,}800.75$ s (about 18:00 hr).

\begin{figure}[H]
\centering
\includegraphics[width=\textwidth]{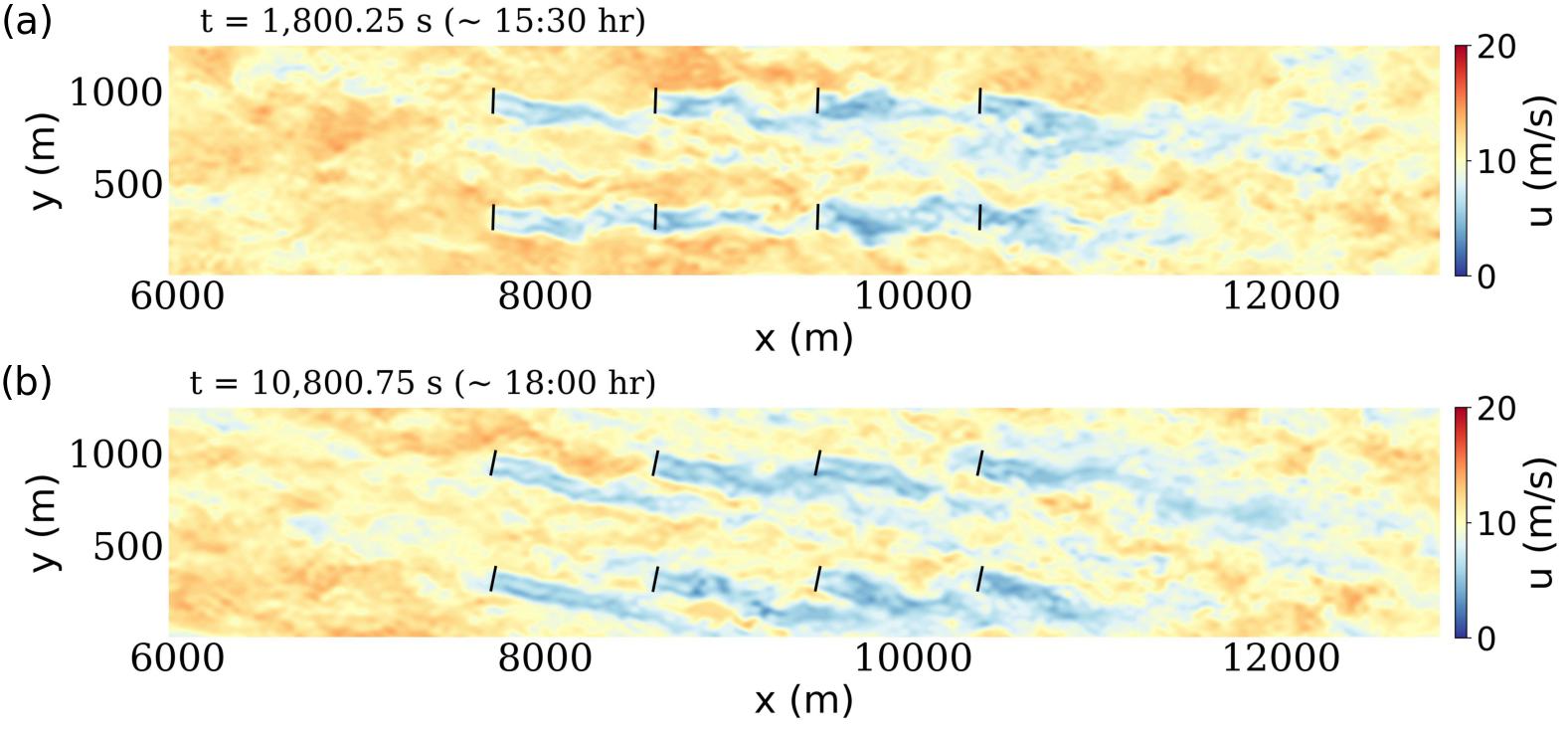}
\caption{Instantaneous streamwise velocity at hub height recorded at (a) $t=1{,}800.25$ s (about 15:30 hr). (b) $t=10{,}800.75$ s (about 18:00 hr). The black solid lines represent the locations of the wind turbines, in which the yaw angle and the locations are obtained from turbine and individual blade data.}
\label{fig:u_diurnal_demo}
\end{figure}

First, coordinates of a set of points are specified (in this case a equally spaced grid of $1{,}100 \times 200$ points in the x and y directions at a constant height $z=z_h=90$m. Note that these points may not coincide with simulation grid points and JHTDB provides appropriately interpolated values \cite{li2008public}. We here specify Lagrange polynomial of order 8 for spatial interpolation. Note that near non-periodic boundaries (bottom and top surfaces, and inflow and outflow planes) interpolation orders are downgraded to accommodate the smaller number of grid points available. For instance, in the last interval between the first and second stored grid-points, only linear interpolation is possible.  The requested times also do not correspond exactly to simulation time steps. JHTDB interpolates in time using third-order CHIP interpolation method \cite{li2008public}. The resulting contour plots are shown in Fig.  \ref{fig:u_diurnal_demo}.

\begin{figure}[H]
\centering
\includegraphics[width=\textwidth]{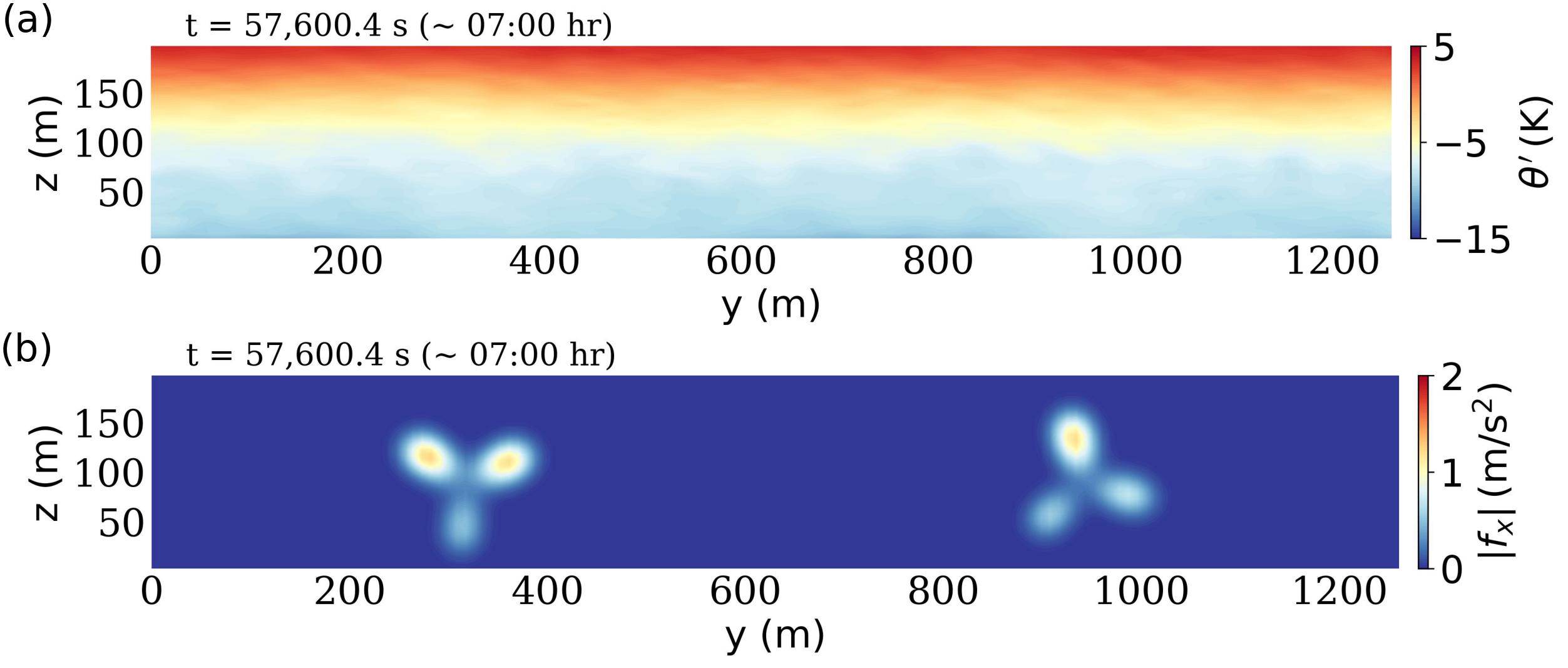}
\caption{(a) Instantaneous temperature field on a y-z plane going through the first row of turbines (turbines 1 and 2, at $x=1{,}764$m downstream of the wind farm domain) recorded at  $t=57{,}600.4$ s (about 07:00 hr). (b) Instantaneous magnitude of x-component of wind turbine force $\left|f_x\right|$ on a y-z plane going through the first row of turbines (turbines \#1 and \#2, at $x=1{,}764$m downstream of the wind farm domain) recorded at  $t=57{,}600.4$ s (about 07:00 hr).}
\label{fig:temp_force_diurnal_demo}
\end{figure}

\begin{figure}[H]
\centering
\includegraphics[width=\textwidth]{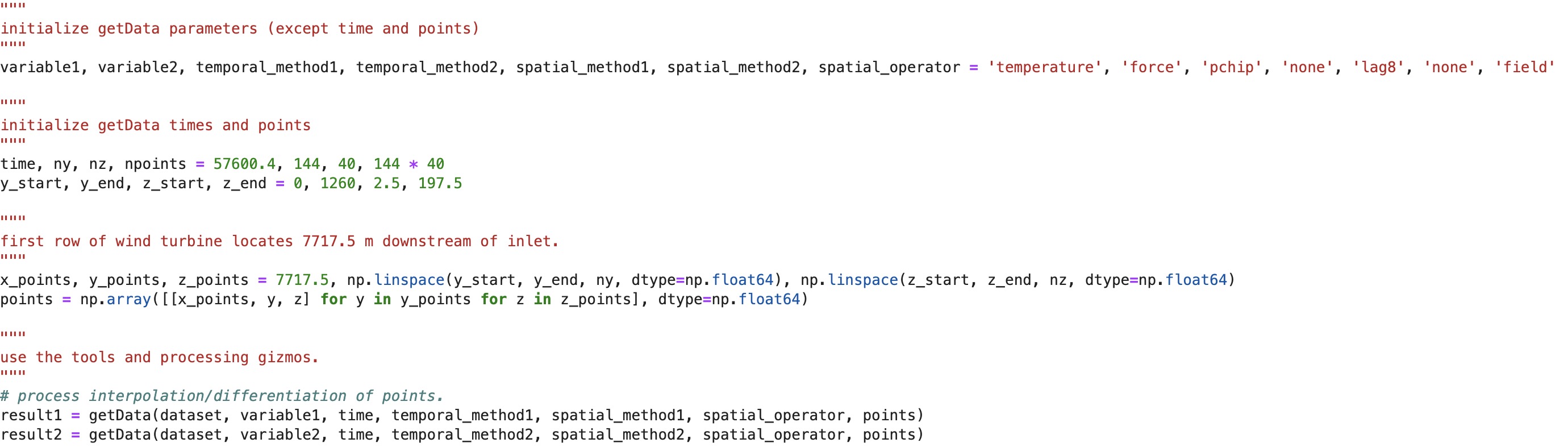}
\caption{Snippets of Python code that queries JHTDB-wind to obtain the temperature field and turbine force-field (its magnitude) distribution on a y-z plane going through the first row of turbines (turbines \#1 and \#2, at time $t=57{,}600.4$s (about 07:00 hr).}
\label{fig:temp_force_diurnal_demo_snippets}
\end{figure}

Similar calls can be made to extract any of the other 3D fields at any of the times. For example, the potential temperature field and turbine force-field (its magnitude) distribution on a y-z plane going through the first row of turbines (turbines \#1 and \#2, at $x=1{,}764$m downstream of the wind farm domain) at time $t=57{,}600.4$s (about 07:00 hr) is shown in Fig. \ref{fig:temp_force_diurnal_demo}. This resulted from code specifications as in the snippet shown in Fig. \ref{fig:temp_force_diurnal_demo_snippets}. More details are provided in \cite{Zhu2025jhtdbwindpaper}.

\begin{figure}[H]
\centering
\includegraphics[width=\textwidth]{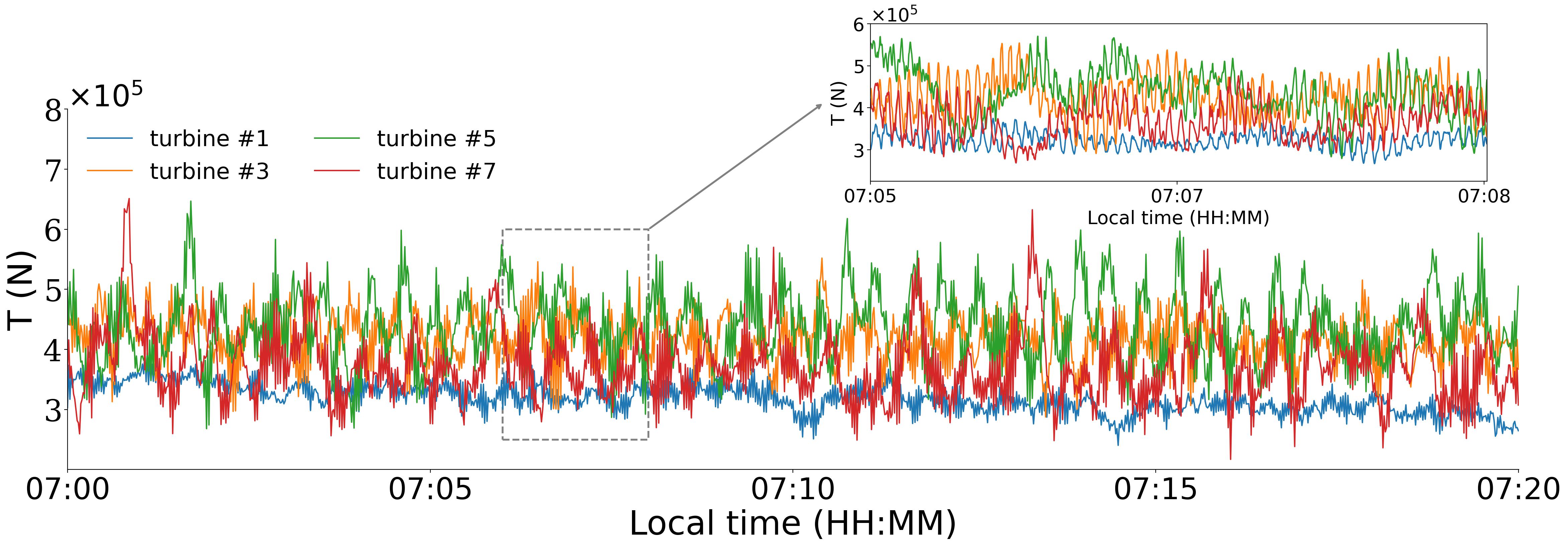}
\caption{Time evolution of the thrust force on the four turbines in a column from 07:00 hr to 08:00 hr.}
\label{fig:thrust_diurnal_demo}
\end{figure}

\begin{figure}[H]
\centering
\includegraphics[width=\textwidth]{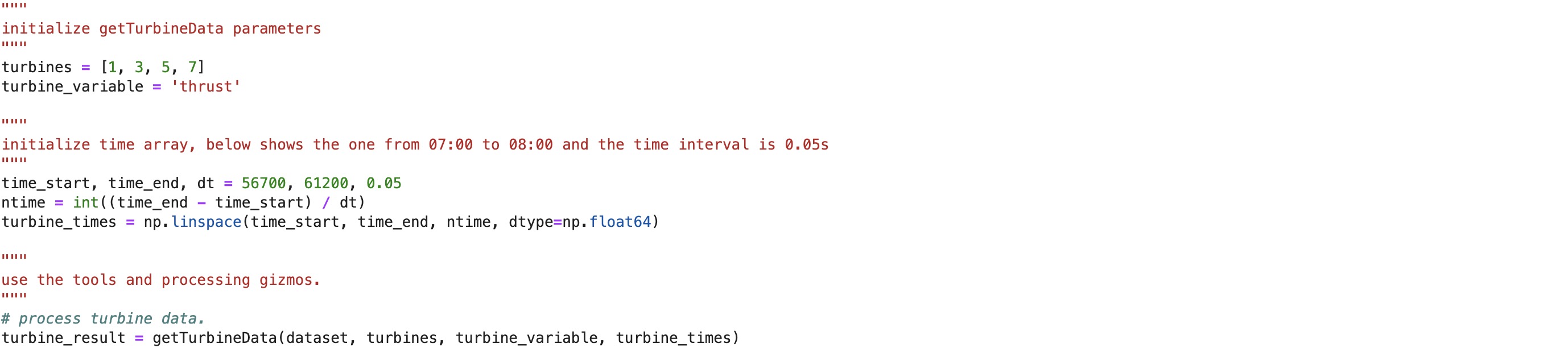}
\caption{Snippets of Python code that queries JHTDB-wind to obtain a time series of thrust force on turbine \#1, \#3, \#5, and \#7, from 07:00 hr to 08:00 hr.}
\label{fig:thrust_diurnal_demo_snippets}
\end{figure}

Turbine and individual blade data are obtained in a similar fashion using get functions, namely {\it getTurbineData(...)} and {\it getBladeData(...)}. Users must specify turbine number for the former, and turbine number and blade number for the latter. An array of point indices along the blade's discretized actuator line is prescribed, as well as an array of times at which the data are required. Linear time interpolation between stored simulation time-steps is supported. More details are provided in \cite{Zhu2025jhtdbwindpaper}. As an example, Fig. \ref{fig:thrust_diurnal_demo} shows a time series of thrust force on four turbines in a column during a one hour interval. Time-delayed travel time correlations can be discerned \cite{wilczek2015spatio}. The code snippet specifying the {\it getTurbineData}(...) call is shown in Fig. 
\ref{fig:thrust_diurnal_demo_snippets}.

\begin{figure}[H]
\centering
\includegraphics[width=\textwidth]{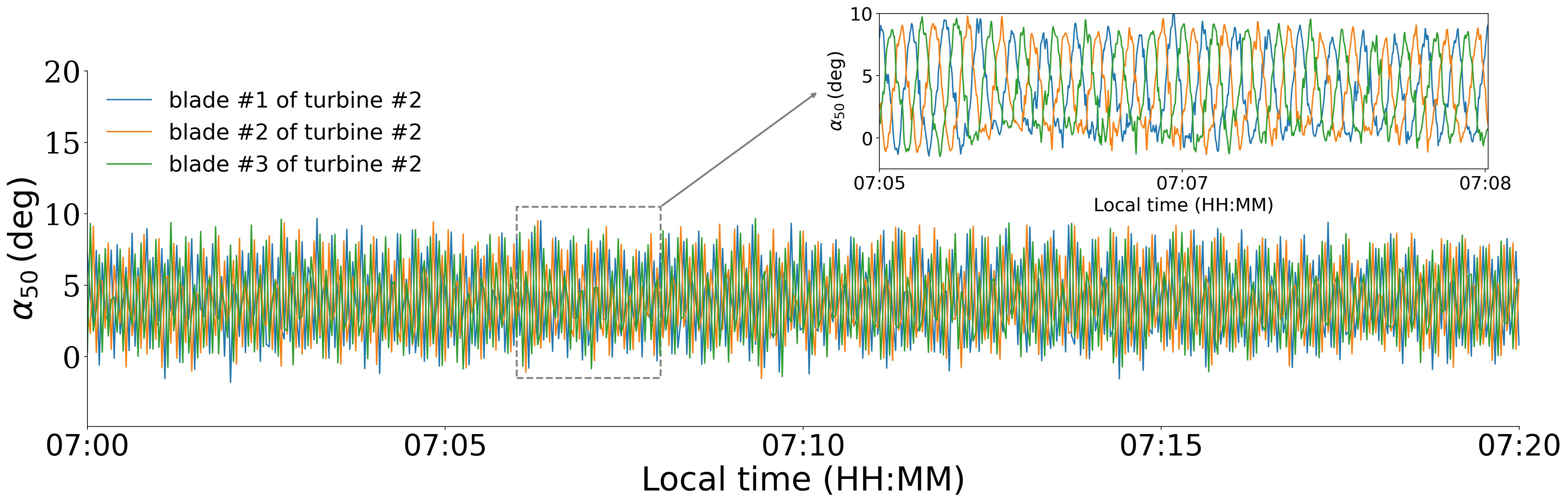}
\caption{Time evolution of the angle of attack on the three blades of turbine \#2 from 07:00 hr to 08:00 hr.}
\label{fig:alpha_diurnal_demo}
\end{figure}

\begin{figure}[H]
\centering
\includegraphics[width=\textwidth]{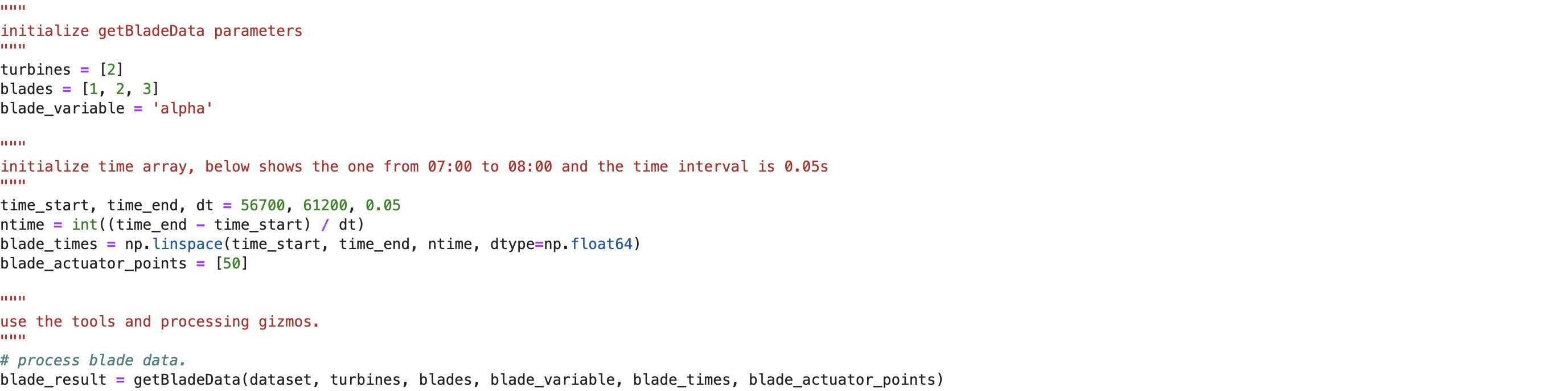}
\caption{Snippets of Python code that queries JHTDB-wind to obtain angle of attack time-histories on blade \#2, from 07:00 hr to 08:00 hr. }
\label{fig:alpha_diurnal_demo_snippets}
\end{figure}

 The use of {\it getBladeData }(...) is illustrated in  Fig. \ref{fig:alpha_diurnal_demo}, showing the angle of attack time-histories on blade \#2. The code snippet specifying the {\it getBladeData}(...) call is shown in Fig. \ref{fig:alpha_diurnal_demo_snippets}
 
\bibliographystyle{unsrt}  
\bibliography{WD}  %%% Remove comment to use the external .bib file (using bibtex).

\end{document}